\documentstyle[amssymb,epsf,graphics,a4,12pt]{article}

\begin{document}

\begin{center}
{\Large\bf Scalar wave propagation in topological black hole backgrounds}
\end{center}

\vspace{1ex} 
\centerline{\large Bin Wang$^{a,b,}$\footnote[1]{e-mail:binwang@fma.if.usp.br},
\ Elcio Abdalla $^{b,}$\footnote[2]{e-mail:eabdalla@fma.if.usp.br}
and R. B. Mann$^{c,}$\footnote[3]{e-mail:rbmann@sciborg.uwaterloo.ca}
}

\begin{center}
{$^{a}$ Department of Physics, Fudan University, Shanghai 200433, P. R.
China \\[0pt]
$^{b}$ Instituto De Fisica, Universidade De Sao Paulo, C.P.66.318, CEP
05315-970, Sao Paulo, Brazil \\[0pt]
$^{c}$ Department of Physics, University of Waterloo, Waterloo, Ontario,
Canada N2L 3G1}
\end{center}

\vspace{6ex}

\vspace{6ex} \hspace*{0mm} PACS number(s): 04.30.Nk, 04.70.Bw 
\begin{abstract}
We consider the evolution of a scalar field coupled to curvature in
topological black hole spacetimes. We solve numerically the scalar wave
equation with different curvature-coupling constant $\xi$ and show that a
rich spectrum of wave propagation is revealed when $\xi$ is introduced.
Relations between quasinormal modes and the size of different topological
black holes have also been investigated.
\end{abstract}

\vfill\eject

\section{Introduction}

Radiative wave propagation in spacetimes containing a black hole has been a
subject of investigation for a number of years. It is well known that after
an initial pulse, the waveform undergoes quasinormal ringing outside the
black hole. The frequencies of the ``ring'' are independent of the processes
which give rise to these oscillations, and instead are directly connected to
the parameters of the black hole. It is generally believed that these
quasinormal frequencies carry a unique fingerprint that leads to direct
identification of a black hole, thereby allowing confirmation of their
existence through anticipated gravitational wave observation in the near
future \cite{kok-sch}.

A great deal of effort has been devoted to the study of the quasinormal
modes associated with black holes. Most of these studies were concerned with
black holes immersed in an asymptotically flat spacetime. The perturbations
of Schwarzschild and Reissner-Nordstrom (RN) black holes can be reduced to
simple wave equations which have been examined extensively \cite{price}\cite%
{g-p-p}\cite{bick}. However, for nonspherical black holes one has to solve
coupled wave equations for the radial part and angular part, respectively.
For this reason the nonspherical case has been studied less thoroughly,
although there has recently been progress along these lines \cite{hod}\cite%
{b-o}\cite{a-g}. In asymptotically flat black hole backgrounds radiative
dynamics always proceeds in the same three stages: initial impulse,
quasinormal ringing and inverse power-law relaxation.

Increasing interest has been expressed in non-asymptotically flat
spacetimes. For a black hole is immersed in an expanding universe, it has
been shown that while the first two stages of radiative dynamics are not
affected by the different conditions at infinity, the third one changes \cite%
{brady1}\cite{brady2}\cite{l-p}. At late times the field decays
exponentially, in contrast to an inverse power law as in the asymptotically
flat case. More thorough exploration of the quasinormal modes of scalar
fields coupled to the curvature have shown that the field's behaviour
strongly depends on the value of the curvature coupling constant.

The first study of the quasinormal ringing for conformally coupled scalar
waves in AdS space was performed by Chan and Mann \cite{mann1}\cite{mann2}.
\ Interest in the asymptotically AdS case has since grown considerably,
motivated by the recent discovery of the AdS/CFT correspondence. In this
context one expects that the quasinormal frequencies of AdS black holes have
a direct interpretation in terms of the dual conformal field theory (CFT) on
the boundary of the asymptotically AdS spacetime. Recently, Horowitz and
Hubeney \cite{horo} considered scalar quasinormal modes on the background of
Schwarzschild AdS black holes in four, five and seven dimensions. They
claimed that for large AdS black holes both the real and imaginary parts of
the quasinormal frequencies scale linearly with the black hole temperature.
However for small AdS black holes they found a departure from this
behaviour. This was further confirmed by the object picture obtained in ref. %
\cite{wang1}. These investigations were further generalized to
Reissner-Nordstrom (RN) AdS black holes, which provide a broader parameter
space than uncharged Schwarzchild AdS (SAdS) black holes \cite{wang2}\cite%
{wang3}. Unlike the SAdS case, the quasinormal frequencies do not scale
linearly with the black hole temperature, and the approach to thermal
equilibrium in the CFT was more rapid as the charge on the black hole
increased. Further extensions in asymptotically AdS spacetimes have included
investigations of quasinormal modes due to electromagnetic and gravitational
perturbations \cite{lemos}, and of nonspherical Kerr-Newmann AdS black hole
backgrounds \cite{ele}.

So far all investigations on wave dynamics in AdS spacetimes have been
limited to fields that are either conformally coupled \cite{mann1}\cite%
{mann2} or minimally coupled to curvature \cite{horo,wang1,wang2,wang3,lemos}%
. In de Sitter spacetimes the behaviour of the scalar field has considerable
dependence on the value of the curvature-coupling constant, and we are
likewise motivated to study the dependence of scalar quasinormal modes on
the curvature-coupling constant in AdS spacetimes, a study we carry out in
this paper. We generalize the study to include topological black holes,
whose quasinormal ringing properties have only been examined in the
conformally coupled case \cite{mann2}. Topological black hole spacetimes are
all asymptotically AdS spacetimes, and have black hole event horizons which
are compact two-surfaces of arbitrary genus \cite{mann3}, with genus zero
being the SAdS case. We solve numerically the wave equations with a general
non-minimal coupling of the scalar field to the spacetime curvature. We find
that the number of possible different behaviours of quasinormal modes for
scalar field is significantly enhanced as the non-minimal coupling parameter 
$\xi $ is varied.

Our paper is organized as follows: in section 2 we review the basic
structures of topological black hole spacetimes and derive the equation
governing the scalar test field coupled to curvature. In the next section we
analyze behaviours of scalar waves in spherical (genus $g=0$), toroidal ($%
g=1 $) and higher genus ($g\geq 2$) black hole backgrounds and their
dependence on curvature coupling constants. Conclusions and a discussion of
resutls are included in the last section.

\section{Equations and numerical methods}

The spacetime of an uncharged topological black hole of arbitrary genus $%
g\geq 0$ has the general form \cite{mann3}\cite{mann1}

\begin{equation}
{\rm d}s^2=-N_g(r){\rm d}t^2+{N_g(r)}^{-1}{\rm d}r^2+r^2{\rm d}\Omega^2,
\end{equation}
\begin{equation}
N_g(r)=r^2/l^2-\epsilon (g-1) - 2M/r
\end{equation}
where 
\begin{equation}
\epsilon(g-1)=[\Theta(g-1)-\Theta(1-g)]=\left\{ 
\begin{array}{ll}
-1 & g=0 \nonumber \\ 
0 & g=1 \\ 
1 & g>1\nonumber%
\end{array}
\right.
\end{equation}
\begin{equation}
{\rm d}\Omega^2=\left\{ 
\begin{array}{ll}
{\rm d}\theta^2+\sin^2\theta{\rm d}\phi^2 & g=0 \nonumber \\ 
{\rm d}\theta^2+{\rm d}\phi^2 & g=1 \\ 
{\rm d}\theta^2+\sinh^2\theta{\rm d}\phi^2 & g>1 \nonumber%
\end{array}
\right.
\end{equation}
and $l=\sqrt{3/\vert\Lambda\vert}$.

When $g=0$ the metric is that of Schwarzschild AdS spacetime. When $g=1$,
the coordinates $\theta $ and $\phi $ describe a flat space and the metric
corresponds to a toroidal spacetime. When $g\geq 2$, the coordinate $(\theta
,\phi )$ are coordinates of a hyperbolic space and the spacetime is referred
to as a topological black hole. Note that the topology of the event horizon
matches the topology out to infinity.

We consider a massless scalar field $\Phi $ in spacetime (1) obeying the
wave equation 
\begin{equation}
(\Box -\xi R)\Phi =0
\end{equation}%
where $\Box =g^{\alpha \beta }\nabla _{\alpha }\nabla _{\beta }$ is the
d'Alembertian operator, $R=-4|\Lambda |$ the Ricci scalar, and $\xi $ a
tunable curvature coupling constant. If we decompose the scalar field
according to 
\begin{equation}
\Phi =\sum \displaystyle\frac{1}{r}\psi (t,r)Y(\theta ,\phi )
\end{equation}%
then each wave function $\psi (r)$ satisfies the equation 
\begin{equation}
-\displaystyle\frac{\partial ^{2}\psi }{\partial t^{2}}+\displaystyle\frac{%
\partial ^{2}\psi }{\partial {X_{g}}^{2}}=V\psi ,
\end{equation}%
where 
\begin{equation}
V=N_{g}[\displaystyle\frac{2}{l^{2}}(1-6\xi )+\displaystyle\frac{2M}{r^{3}}]
\end{equation}%
and $X_{g}=\int {N_{g}}^{-1}dr$ is the tortoise coordinate.

To compare the wave behaviour for black holes of different topologies, we
follow ref. \cite{mann1} and perform a rescaling $r=r_{+}z$ so that the
event horizon is at unit dimensionless distance. Adopting this dimensionless
variable $z$, we obtain 
\begin{equation}
{\tilde{N}_{g}}=l^{2}N_{g}/r_{+}^{2}=\displaystyle\frac{z-1}{z}(z^{2}+z+1-%
\displaystyle\frac{l^{2}}{r_{+}^{2}}\epsilon (g-1)).
\end{equation}%
After rescaling the scalar wave equation (7) becomes 
\begin{equation}
-\displaystyle\frac{\partial ^{2}{\tilde{\psi}}}{\partial {\tilde{t}}^{2}}+%
\displaystyle\frac{\partial ^{2}{\tilde{\psi}}}{\partial {\tilde{X_{g}}}^{2}}%
={\tilde{V}}{\tilde{\psi}},
\end{equation}%
where 
\begin{eqnarray}
{\tilde{t}} &=&\displaystyle\frac{r_{+}}{l^{2}}t  \label{eq11} \\
{\tilde{X}_{g}} &=&\displaystyle\frac{r_{+}}{l^{2}}X_{g}  \label{eq12} \\
{\tilde{V}} &=&[1-3\beta ^{2}\epsilon (g-1)]\displaystyle\frac{z-1}{z^{4}}%
[z^{2}+z+1-3\beta ^{2}\epsilon (g-1)]  \nonumber \\
&&+2\chi \displaystyle\frac{z-1}{z}[z^{2}+z+1-3\beta ^{2}\epsilon
(g-1)]\quad {\rm and}\quad \chi =1-6\xi  \label{Vhat} \\
\beta &=&\displaystyle\frac{l}{\sqrt{3}r_{+}}>0  \label{eq14}
\end{eqnarray}

For different black holes with different topologies, the rescaled tortoise
coordinates ${\tilde{X}_{g}}$ are \cite{mann1} 
\begin{eqnarray}
{\tilde{X}_{g=0}} &=&\displaystyle\frac{1}{3(1+\beta ^{2})}[\ln \displaystyle%
\frac{z-1}{\sqrt{z^{2}+1+1+3\beta ^{2}}}+\displaystyle\frac{\sqrt{3}%
(1+2\beta ^{2})}{\sqrt{1+4\beta ^{2}}}\arctan (\displaystyle\frac{2z+1}{%
\sqrt{3+12\beta ^{2}}})]  \nonumber \\
{\tilde{X}_{g=1}} &=&\displaystyle\frac{1}{3}\ln \displaystyle\frac{z-1}{%
\sqrt{z^{2}+z+1}}+\displaystyle\frac{1}{3}\arctan (\displaystyle\frac{2z+1}{%
\sqrt{3}})  \nonumber \\
{\tilde{X}_{g\geq 2}} &=&\displaystyle\frac{1}{3(1-\beta ^{2})}[\ln %
\displaystyle\frac{z-1}{\sqrt{z^{2}+1+1-3\beta ^{2}}}+\displaystyle\frac{%
\sqrt{3}(1-2\beta ^{2})}{\sqrt{1-4\beta ^{2}}}\arctan (\displaystyle\frac{%
2z+1}{\sqrt{3-12\beta ^{2}}})]\quad (0<\beta <1/2)  \nonumber \\
{\tilde{X}_{g\geq 2}} &=&\displaystyle\frac{4}{9}\ln \displaystyle\frac{z-1}{%
z+1/2}-\displaystyle\frac{2}{3(2z+1)}\quad (\beta =1/2)  \nonumber \\
{\tilde{X}_{g\geq 2}} &=&\displaystyle\frac{1}{3(1-\beta ^{2})}[\ln %
\displaystyle\frac{z-1}{\sqrt{z^{2}+1+1-3\beta ^{2}}}  \nonumber \\
&&+\displaystyle\frac{\sqrt{3}(1-2\beta ^{2})}{\sqrt{4\beta ^{2}-1}}\ln %
\displaystyle\frac{2z+1-\sqrt{3}\sqrt{4\beta ^{2}-1}}{2z+1+\sqrt{3}\sqrt{%
4\beta ^{2}-1}}]\quad (1/2<\beta <1)  \nonumber \\
{\tilde{X}_{g\geq 2}} &=&\displaystyle\frac{2}{9}\ln \displaystyle\frac{z-1}{%
z+2}-\displaystyle\frac{1}{3(z-1)}\quad (\beta =1)
\end{eqnarray}%
We see that for the toroidal topology the rescaled tortoise coordinate is
independent of the parameter $\beta $. For the higher genus cases, $\beta $
has a range $0<\beta \leq 1$. $\beta >1$ corresponds to the naked
singularity case \cite{mann1}. The black hole mass can be expressed as $%
\sqrt{3}(M/l)=1/\beta ^{3}-3/\beta $ for $g\geq 2$ case. There are two
qualitatively different parts of $\beta $, namely $0<\beta <1/\sqrt{3}$ and $%
1/\sqrt{3}<\beta <1$, which correspond to two different topological black
holes with positive mass and negative mass \cite{mann4}, respectively. $%
\beta =1/\sqrt{3}$ corresponds to the zero mass topological black hole.

Using the null coordinates $u={\tilde{t}}-{\tilde{X}_{g}}$ and $v={\tilde{t}}%
+{\tilde{X}_{g}}$, (10) can be recast as 
\begin{equation}
-4\displaystyle\frac{\partial ^{2}}{\partial u\partial v}{\tilde{\psi}}(u,v)=%
{\tilde{V}}(z){\tilde{\psi}}(u,v)  \label{eq16}
\end{equation}%
in which $z$ is determined by inverting the relation ${\tilde{X}_{g}}%
(z)=(v-u)/2$.

The two-dimensional wave equation (16) can be integrated numerically, using
for example the finite difference method suggested in \cite{g-p-p}. Using
Taylor's theorem, it is discretized as 
\begin{equation}
{\tilde{\psi}}_{N}={\tilde{\psi}}_{E}+{\tilde{\psi}}_{W}-{\tilde{\psi}}%
_{S}-\delta u\delta v{\tilde{V}}(\displaystyle\frac{v_{N}+v_{W}-u_{N}-u_{E}}{%
4})\displaystyle\frac{{\tilde{\psi}}_{W}+{\tilde{\psi}}_{E}}{8}+O(\epsilon
^{4})
\end{equation}%
where the points $N,S,E$ and $W$ form a null rectangle with relative
positions as: $N:(u+\delta u,v+\delta v),W:(u+\delta u,v),E:(u,v+\delta v)$
and $S:(u,v)$. $\epsilon $ is an overall grid scalar factor, so that $\delta
u\sim \epsilon \sim \delta v$. Considering that the behaviour of the wave
function is not sensitive to the choice of initial data, we set ${\tilde{\psi%
}}(u,v=v_{0})=0$ and use a Gaussian pulse as an initial perturbation,
centered on $v_{c}$ and with width $\sigma $ on $u=u_{0}$ as 
\begin{equation}
{\tilde{\psi}}(u=u_{0},v)=\exp [-\displaystyle\frac{(v-v_{c})^{2}}{2\sigma
^{2}}]
\end{equation}%
The inversion of the relation ${\tilde{X}_{g}}(z)$ needed in the evaluation
of the potential ${\tilde{V}}(z)$ is the most tedious part of the
computation.

After the integration is completed, the value ${\tilde{\psi}}(u_{max},v)$ is
extracted, where $u_{max}$ is the maximum value of $u$ on the numerical
grid. Taking sufficiently large $u_{max}$, ${\tilde{\psi}}(u_{max},v)$
represents a good approximation for the wave function at the event horizon.
Since it has been shown that wave behaviour is the same near or far from the
event horizon \cite{g-p-p,mann2}, we will study the dependence of ${\tilde{%
\psi}}(u_{max},v)$ on the genus $g$ non-minimal coupling constant $\xi $ .

\begin{figure}[tbh]
\begin{center}
\leavevmode       
\begin{eqnarray}
\epsfxsize= 8truecm\rotatebox{-90}{\epsfbox{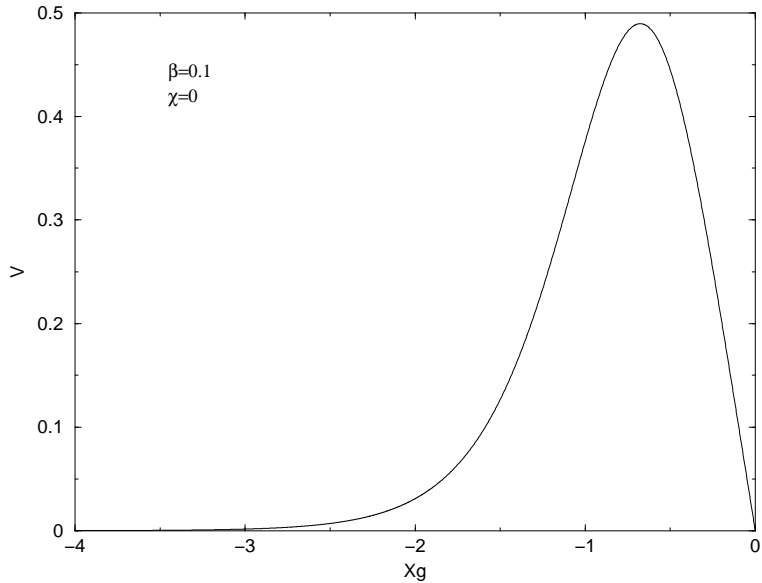}}  \nonumber
\end{eqnarray}
\vskip .5cm
\end{center}
\caption{{Potential behavior for $\protect\xi =1/6$}}
\label{fig1}
\end{figure}

\begin{figure}[tbh]
\begin{center}
\leavevmode       
\begin{eqnarray}
\epsfxsize= 8truecm\rotatebox{-90}{\epsfbox{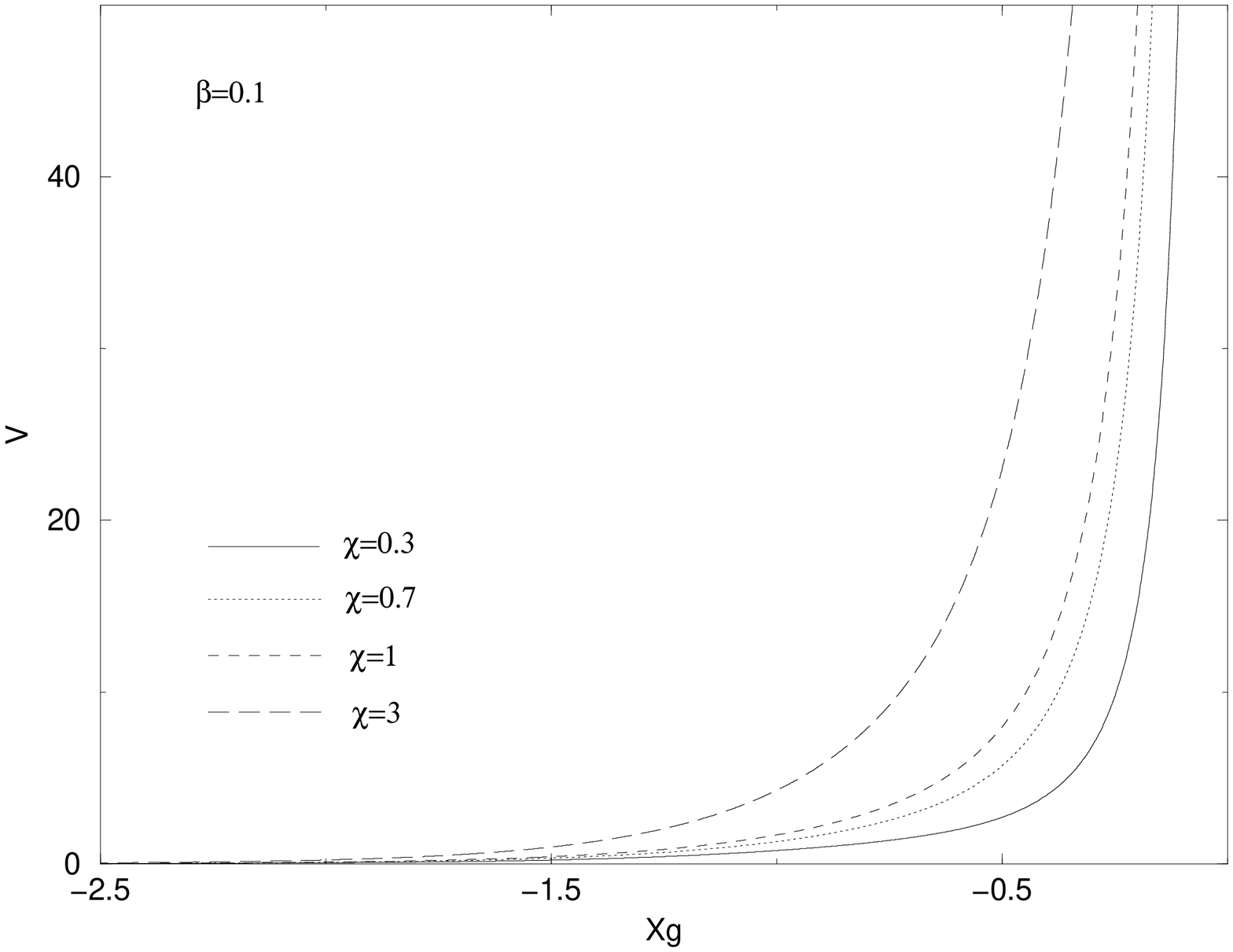}}  \nonumber
\end{eqnarray}
\vskip .5cm
\end{center}
\caption{{Potential behavior for $\protect\xi <1/6$}}
\label{fig2}
\end{figure}

\begin{figure}[tbh]
\begin{center}
\leavevmode       
\begin{eqnarray}
\epsfxsize= 8truecm\rotatebox{-90}{\epsfbox{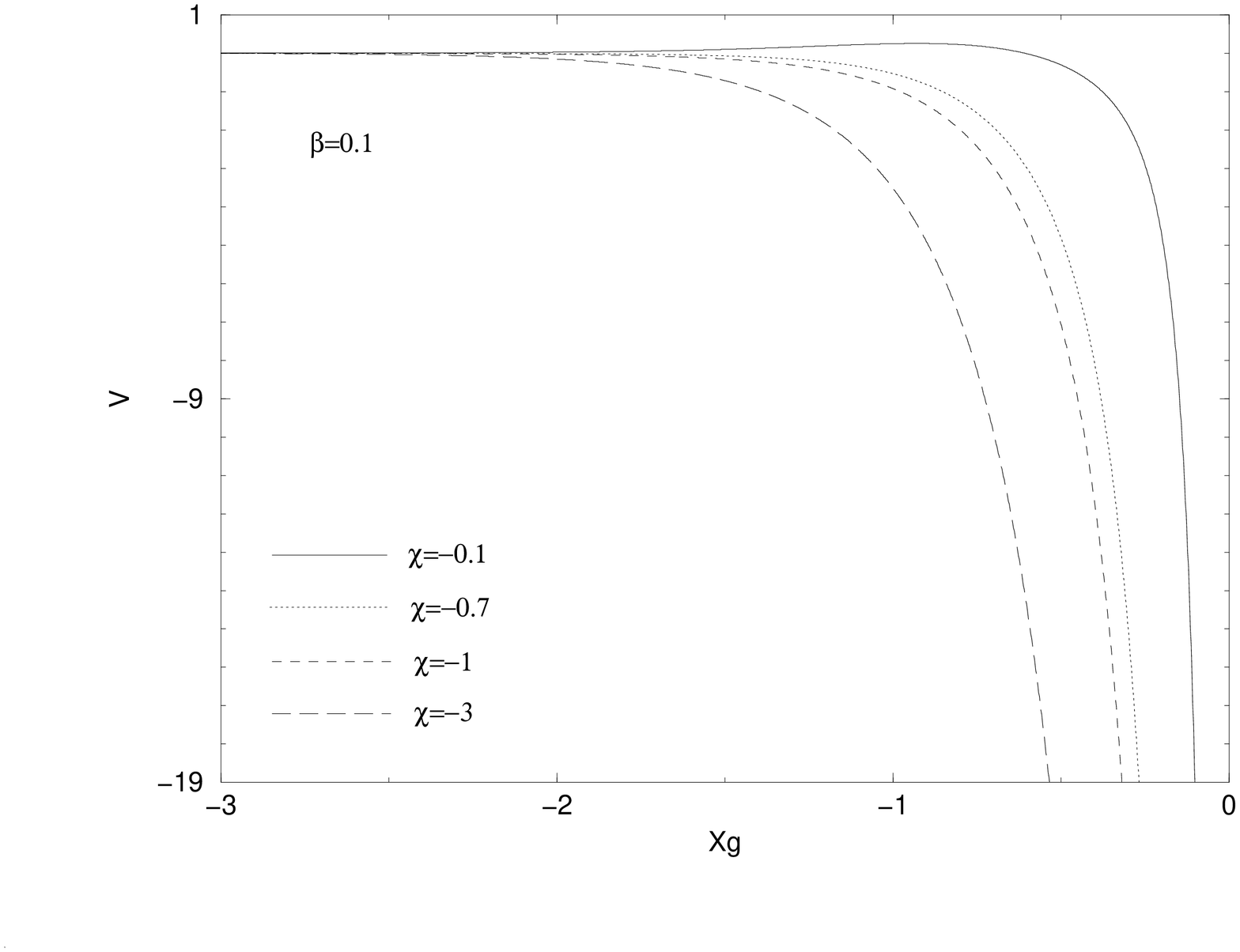}}  \nonumber
\end{eqnarray}
\vskip .5cm
\end{center}
\caption{{Potential behavior for $\protect\xi >1/6$}}
\label{fig3}
\end{figure}

\begin{figure}[tbh]
\begin{center}
\leavevmode       
\begin{eqnarray}
\epsfxsize= 6truecm\rotatebox{-90}{\epsfbox{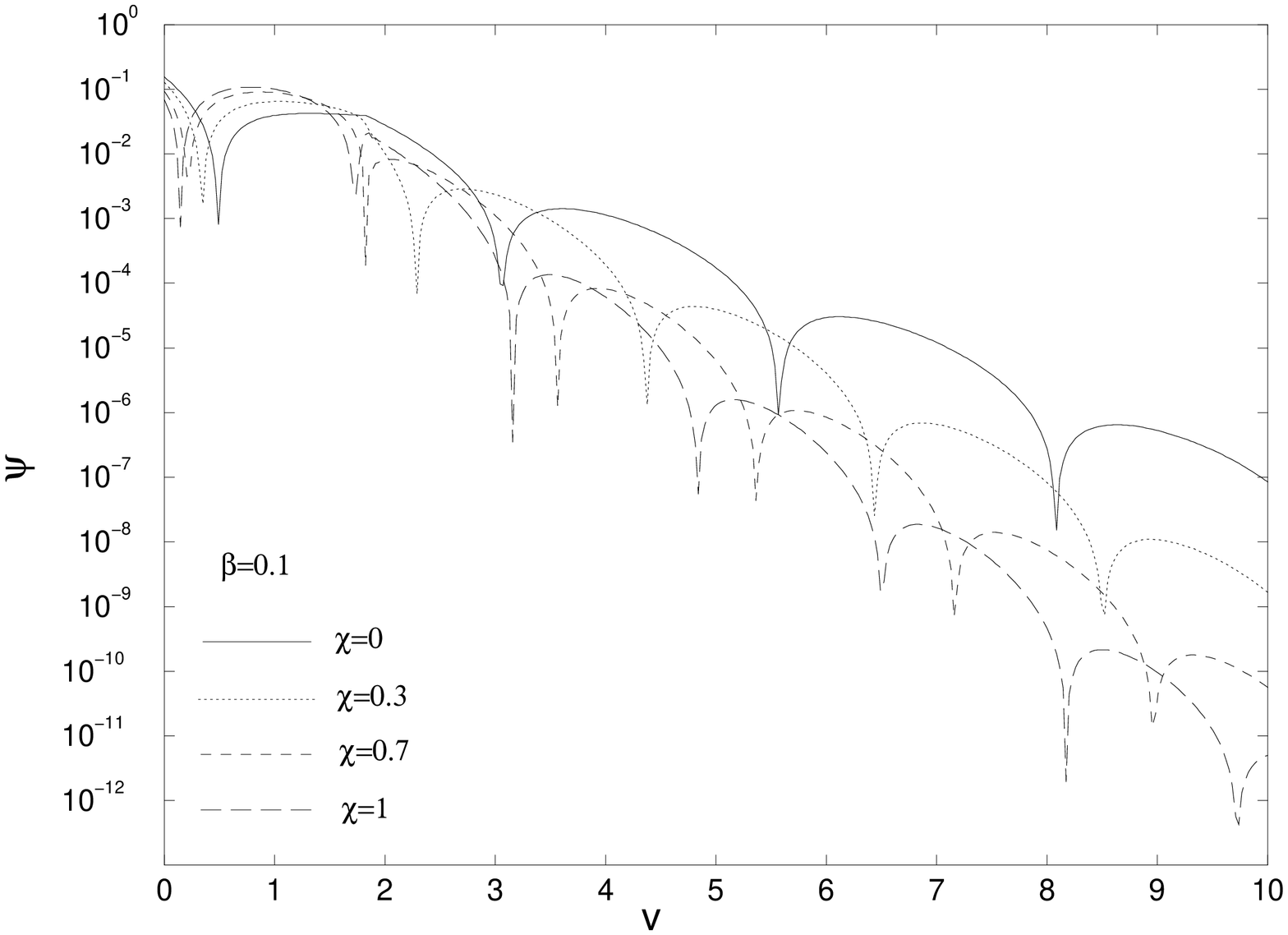}} & & \epsfxsize=6truecm%
\rotatebox{-90}{\epsfbox{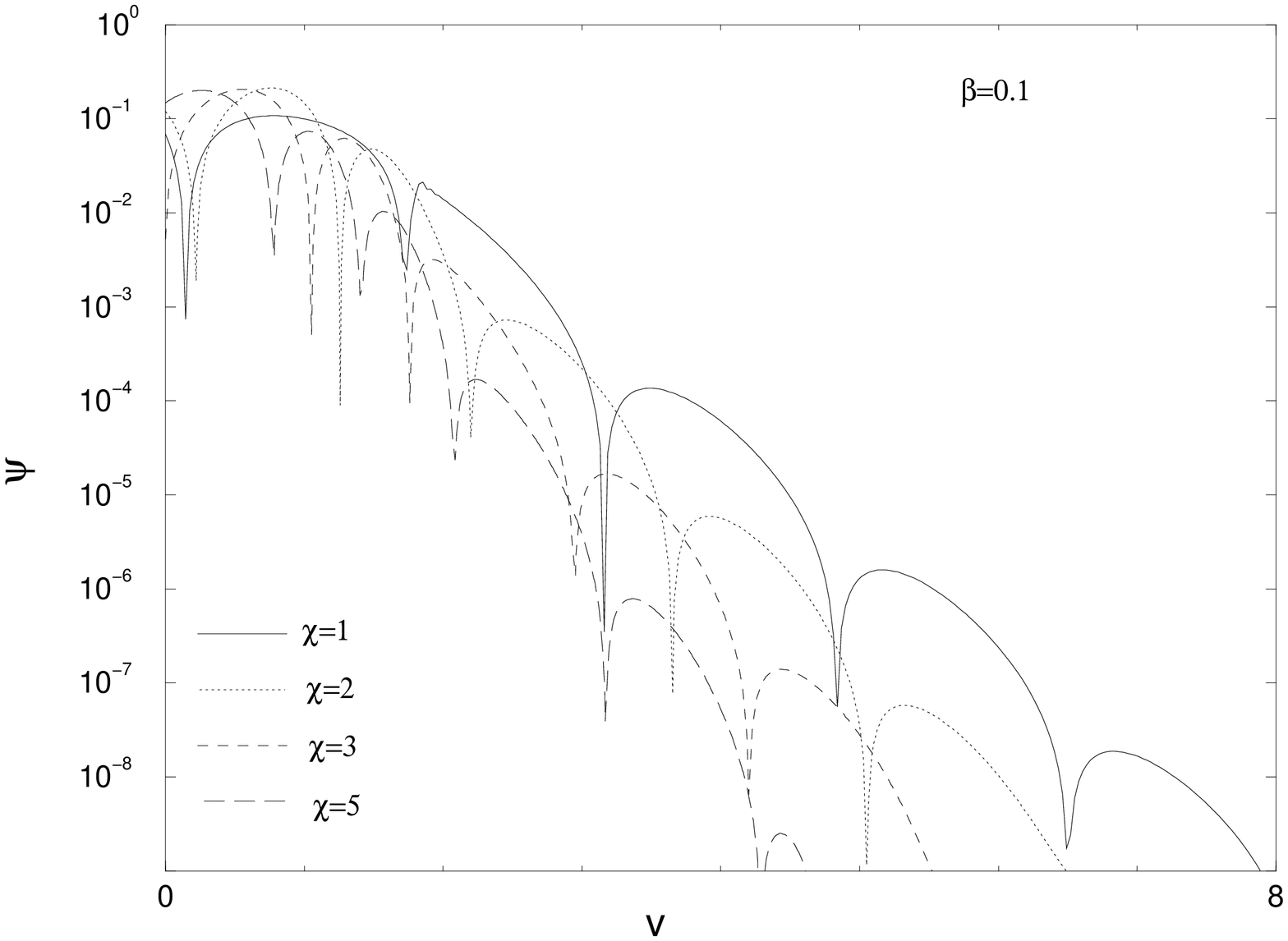}}  \nonumber
\end{eqnarray}
\vskip .5cm
\end{center}
\caption{{Wave propagation of scalar field for $\protect\xi \leq 1/6$ with $%
\protect\beta =0.1$}}
\label{fig4}
\end{figure}
\begin{figure}[tbh]
\begin{center}
\leavevmode       
\begin{eqnarray}
\epsfxsize= 8truecm\rotatebox{-90}{\epsfbox{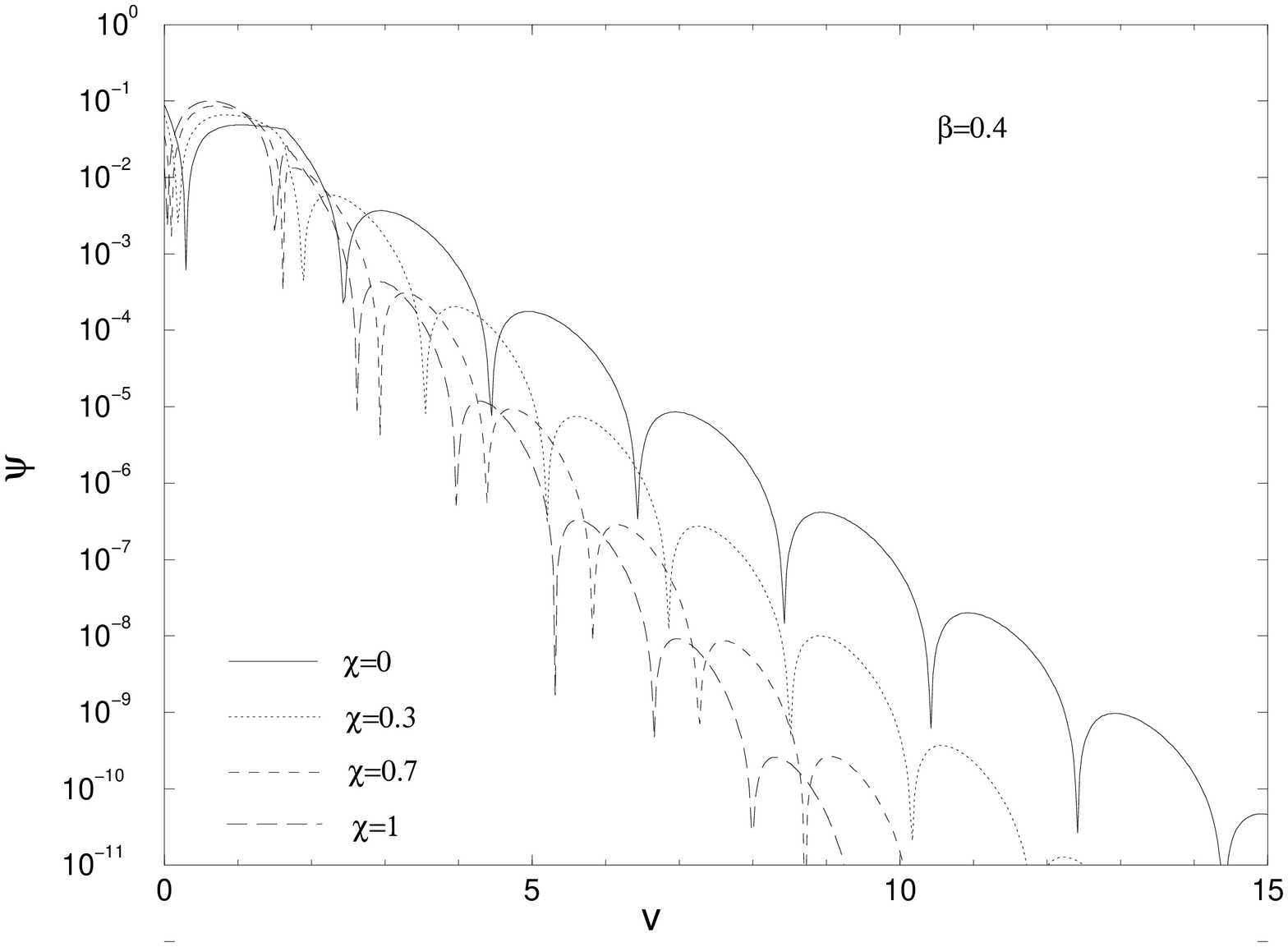}}  \nonumber
\end{eqnarray}
\vskip .5cm
\end{center}
\caption{{\ Wave propagation of scalar field for $\protect\xi \leq 1/6$ with 
$\protect\beta =0.4$}}
\label{fig5}
\end{figure}

\begin{figure}[tbh]
\begin{center}
\leavevmode       
\begin{eqnarray}
\epsfxsize= 8truecm\rotatebox{-90}{\epsfbox{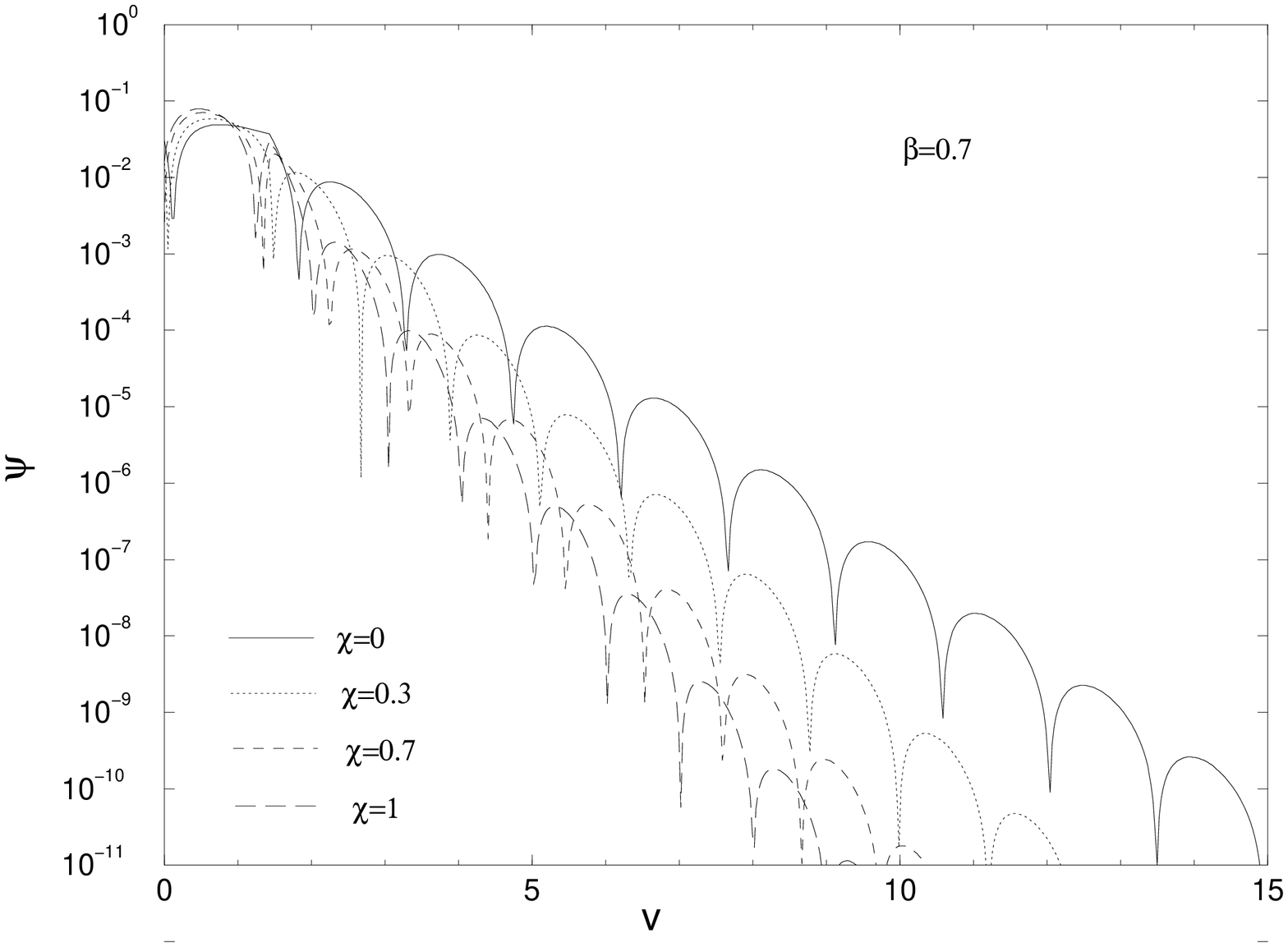}}  \nonumber
\end{eqnarray}
\vskip .5cm
\end{center}
\caption{{\ Wave propagation of scalar field for $\protect\xi \leq 1/6$ with 
$\protect\beta =0.7$}}
\label{fig6}
\end{figure}

\begin{figure}[tbh]
\begin{center}
\leavevmode       
\begin{eqnarray}
\epsfxsize= 6truecm\rotatebox{-90}{\epsfbox{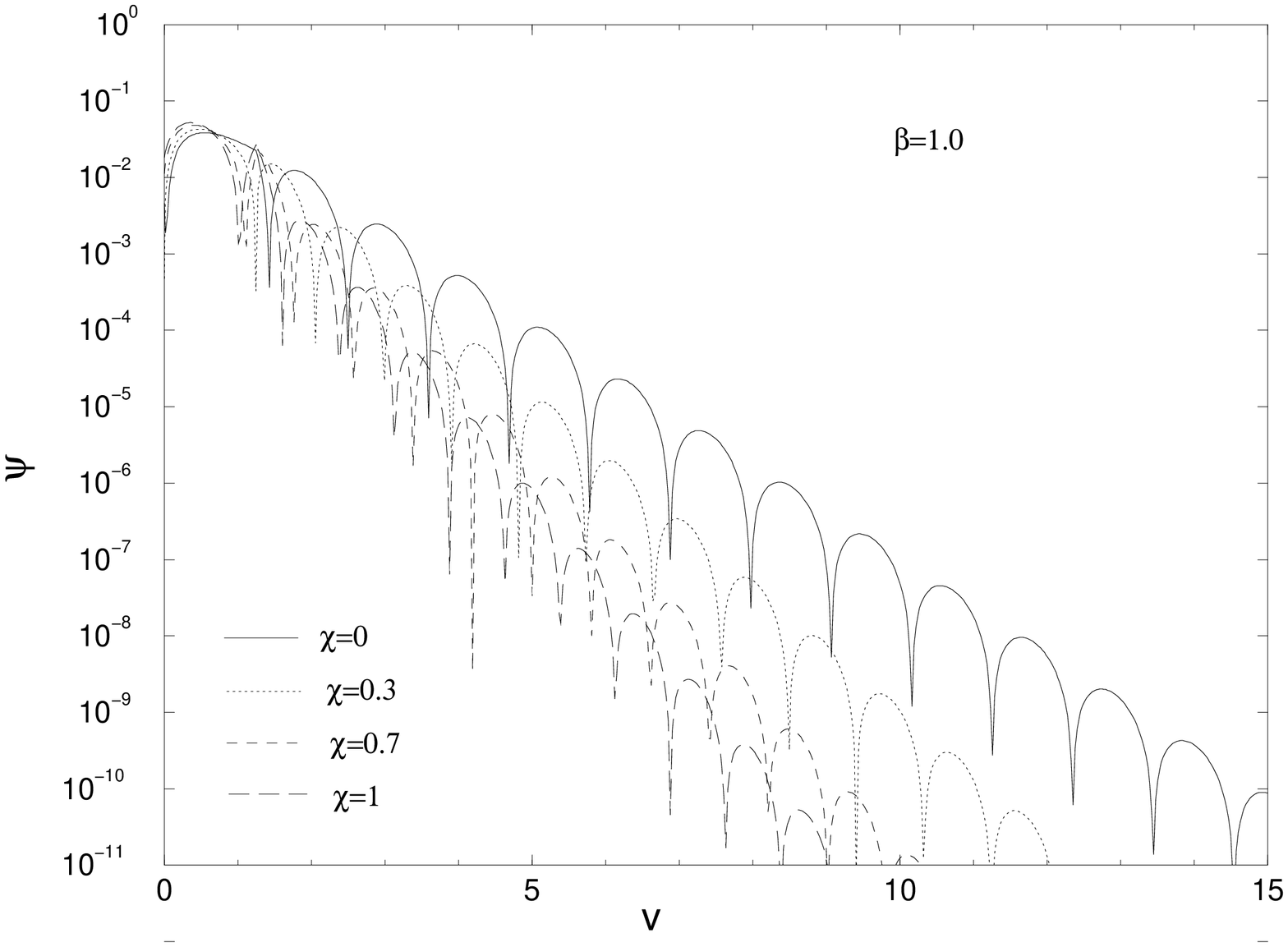}} & & \epsfxsize=6truecm%
\rotatebox{-90}{\epsfbox{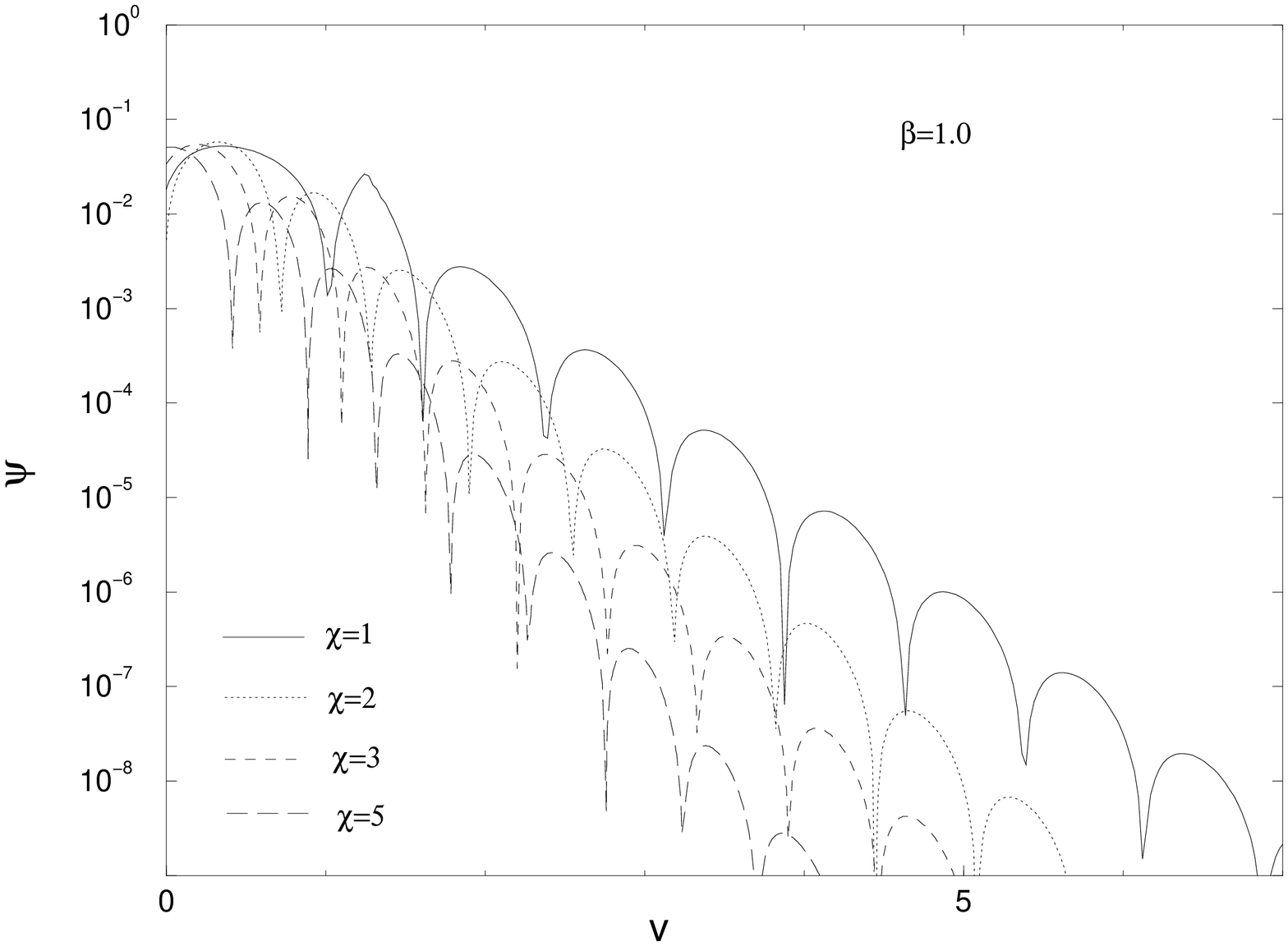}}  \nonumber
\end{eqnarray}
\vskip .5cm
\end{center}
\caption{{Wave propagation of scalar field for $\protect\xi \leq 1/6$ with $%
\protect\beta =1.0$}}
\label{fig7}
\end{figure}
\begin{figure}[tbh]
\begin{center}
\leavevmode       
\begin{eqnarray}
\epsfxsize= 8truecm\rotatebox{-90}{\epsfbox{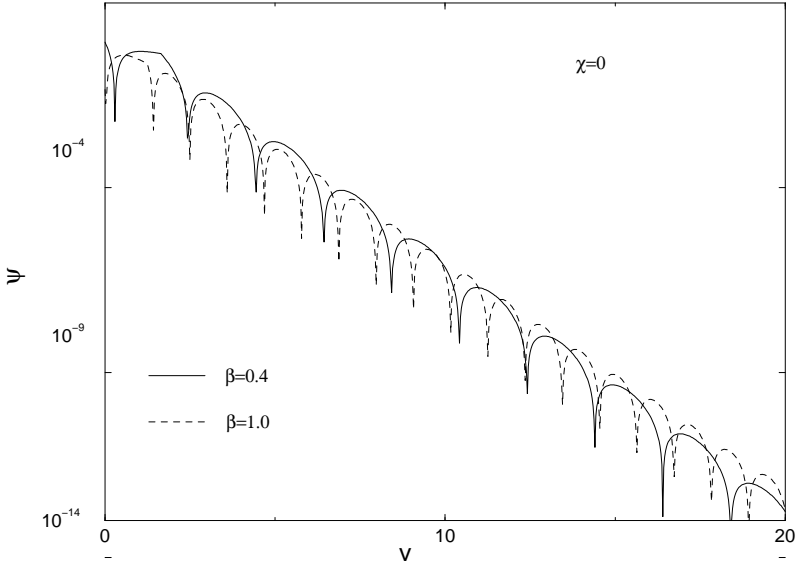}}  \nonumber
\end{eqnarray}
\vskip .5cm
\end{center}
\caption{{\ Behavior of field with the increase of $\protect\beta $}}
\label{fig8}
\end{figure}
\begin{figure}[tbh]
\begin{center}
\leavevmode       
\begin{eqnarray}
\epsfxsize= 6truecm\rotatebox{-90}{\epsfbox{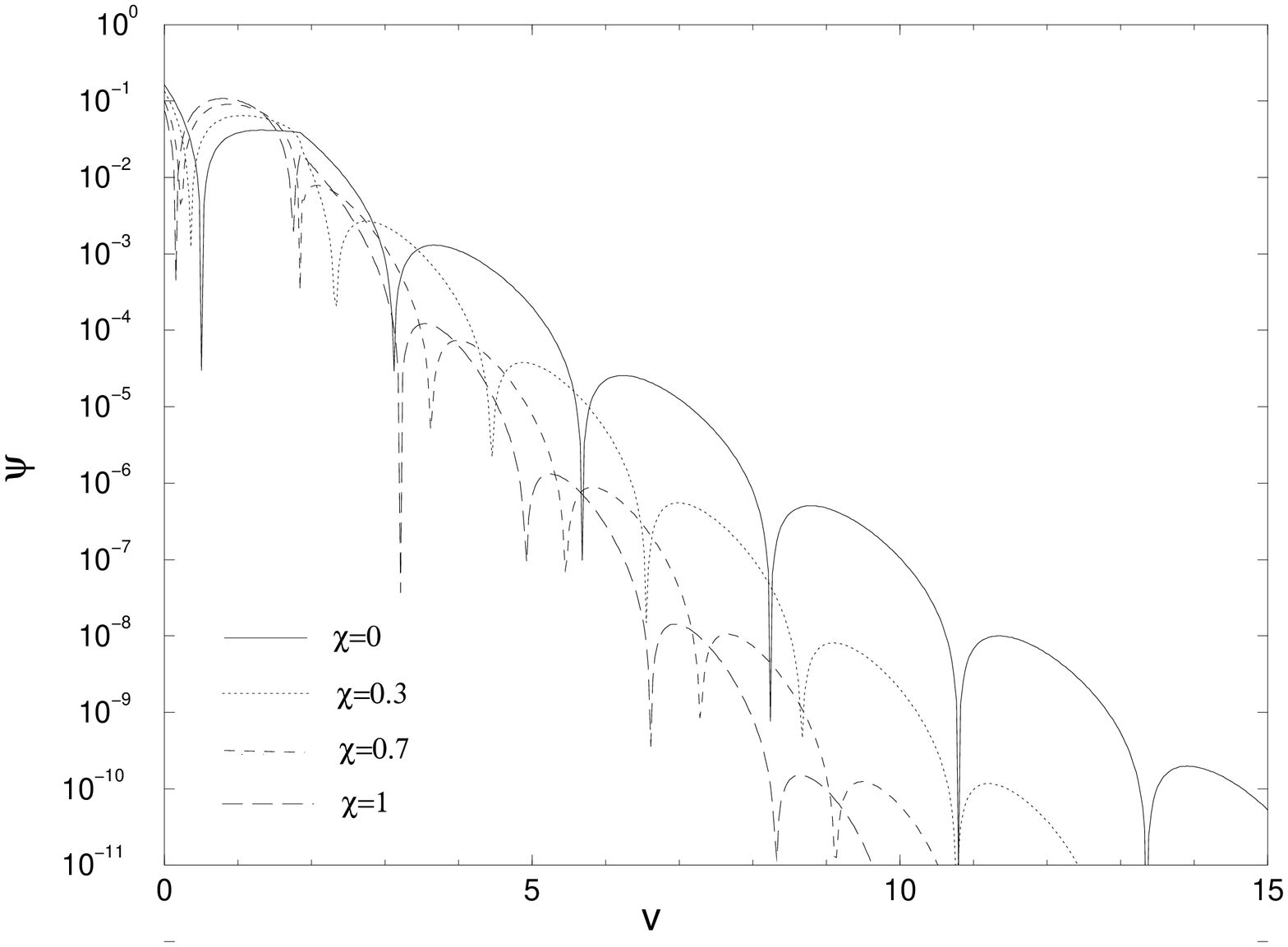}} & & \epsfxsize=6truecm%
\rotatebox{-90}{\epsfbox{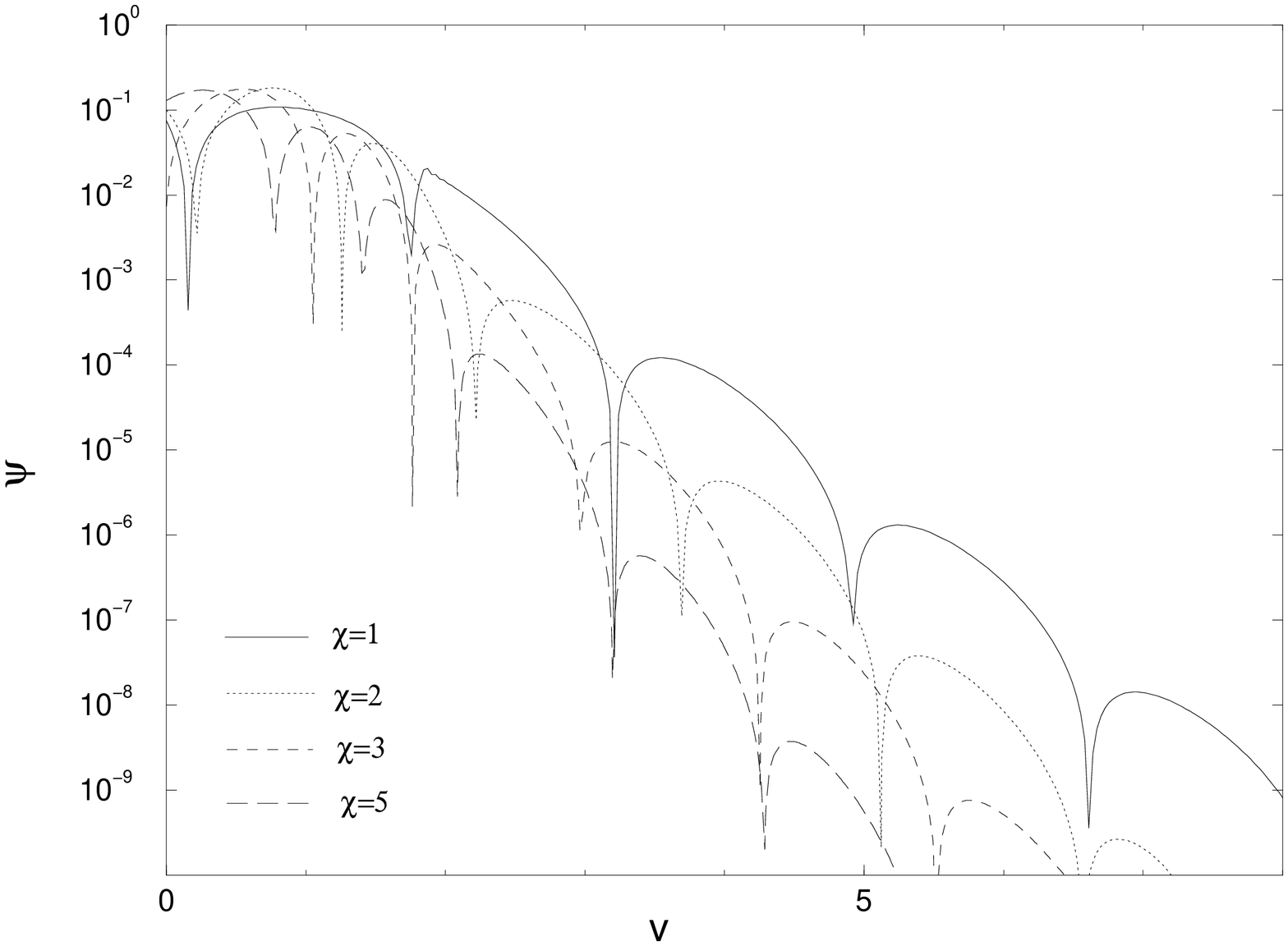}}  \nonumber
\end{eqnarray}
\vskip .5cm
\end{center}
\caption{{Falloff behavior for $\protect\xi \leq 1/6$ in toroidal black holes%
} }
\label{fig9}
\end{figure}
\begin{figure}[tbh]
\begin{center}
\leavevmode       
\begin{eqnarray}
\epsfxsize= 6truecm\rotatebox{-90}{\epsfbox{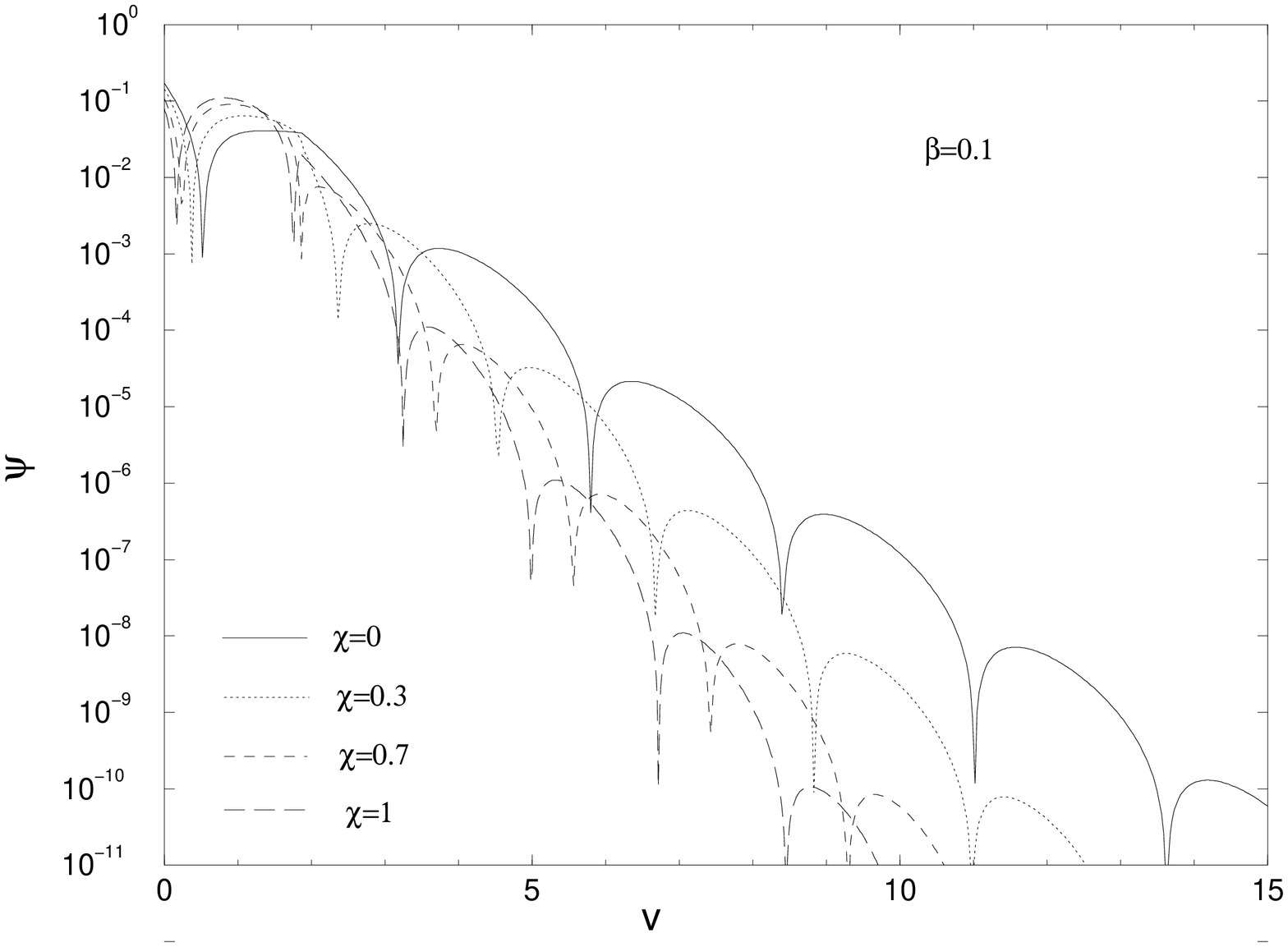}} & & \epsfxsize%
=6truecm\rotatebox{-90}{\epsfbox{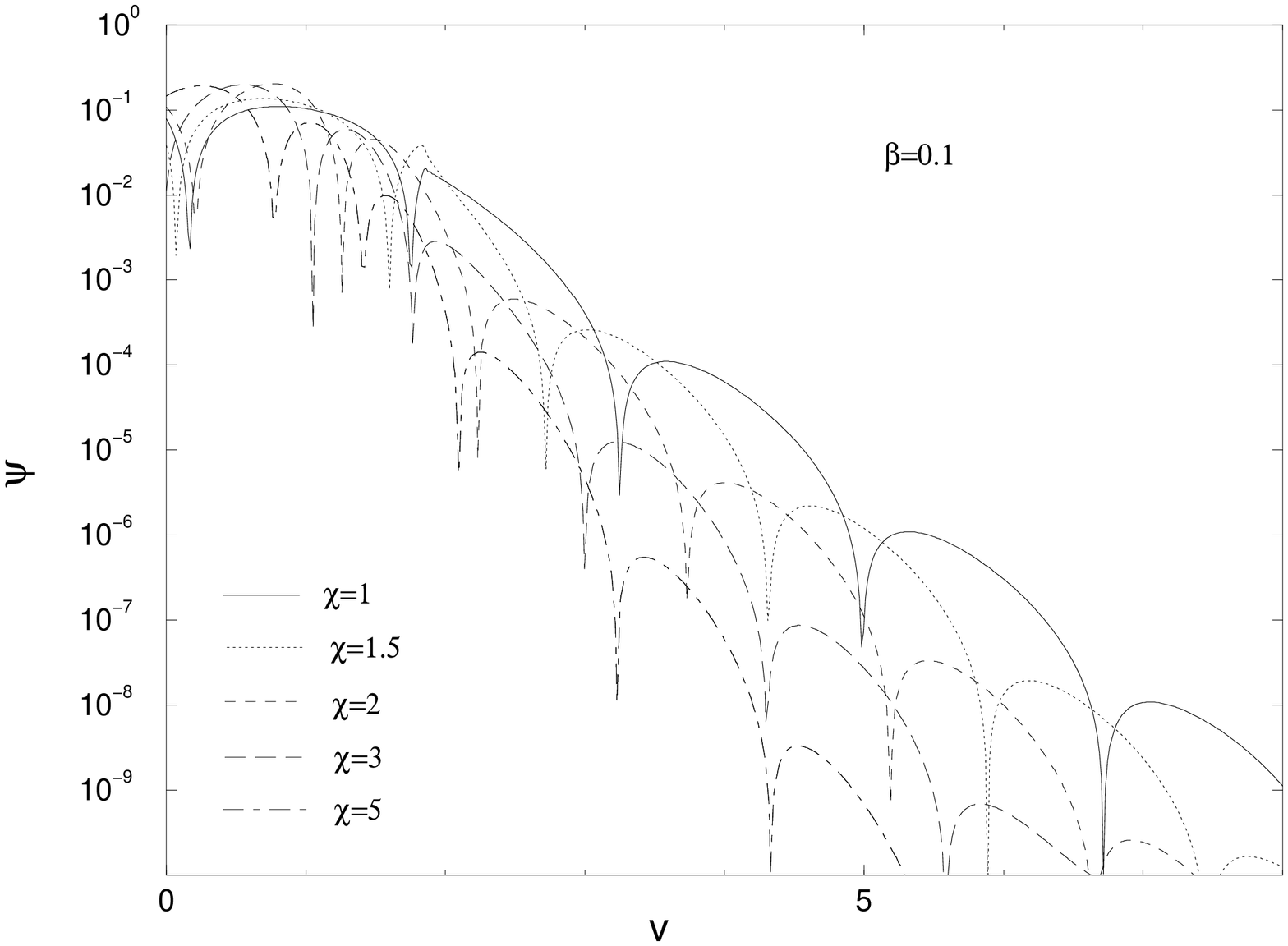}}  \nonumber
\end{eqnarray}
\vskip .5cm
\end{center}
\caption{{Falloff behavior for $\protect\xi \leq 1/6$ in positive mass
topology black holes when $\protect\beta =0.1$}}
\label{fig10}
\end{figure}
\begin{figure}[tbh]
\begin{center}
\leavevmode       
\begin{eqnarray}
\epsfxsize= 6truecm\rotatebox{-90}{\epsfbox{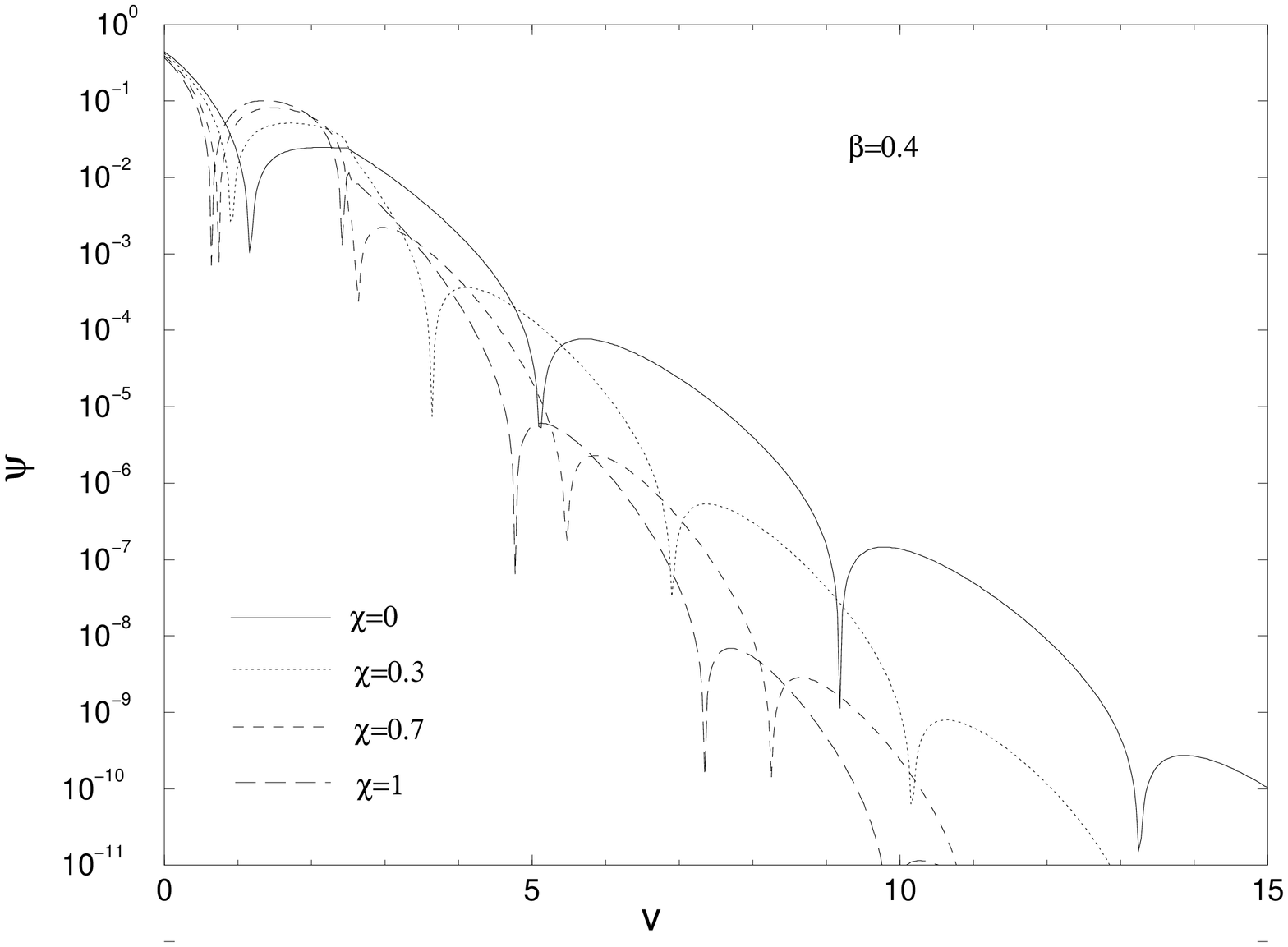}} & & \epsfxsize%
=6truecm\rotatebox{-90}{\epsfbox{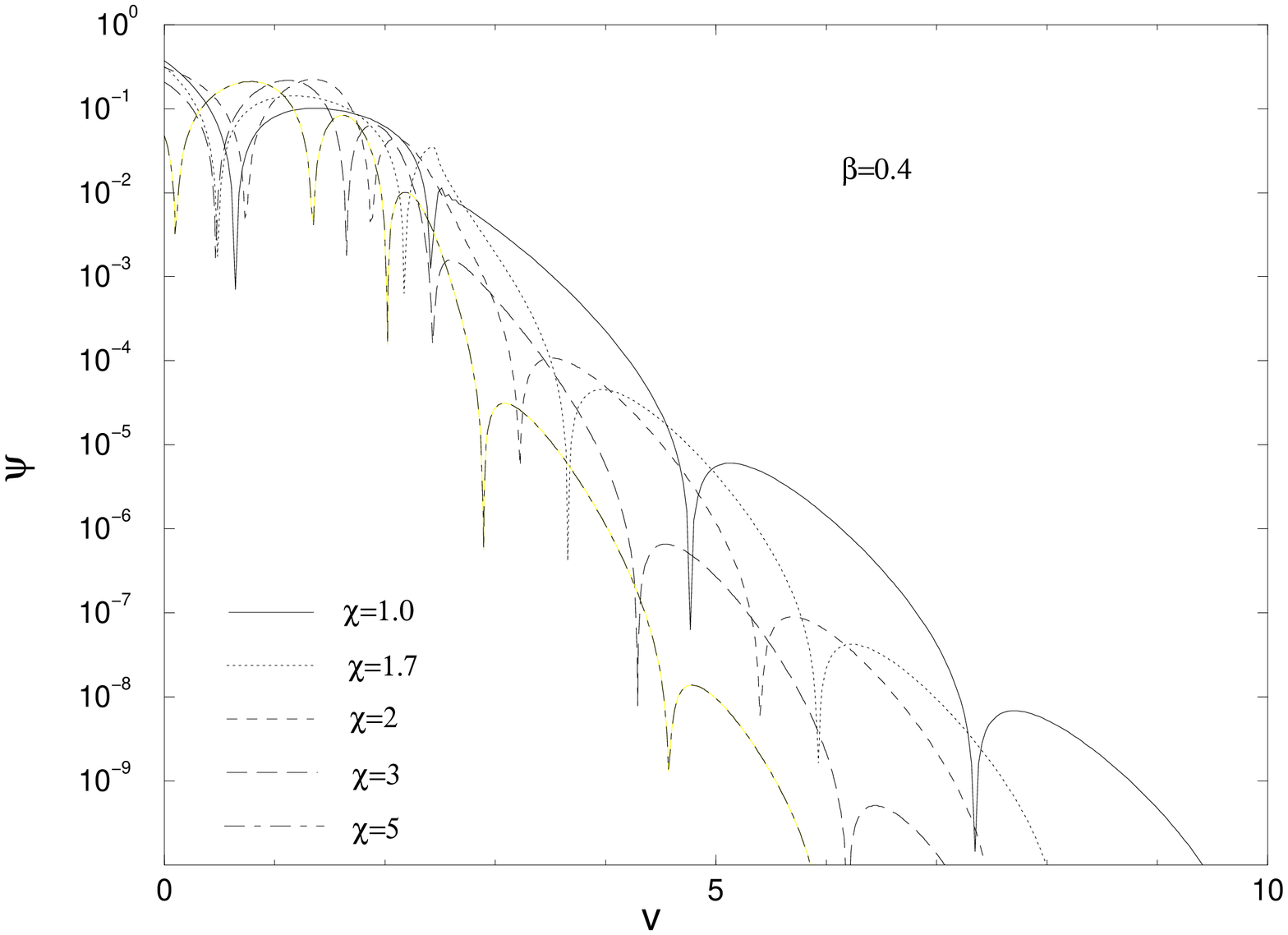}}  \nonumber
\end{eqnarray}
\vskip .5cm
\end{center}
\caption{{Falloff behavior for $\protect\xi \leq 1/6$ in positive mass
topology black holes when $\protect\beta =0.4$}}
\label{fig11}
\end{figure}
\begin{figure}[tbh]
\begin{center}
\leavevmode       
\begin{eqnarray}
\epsfxsize= 6truecm\rotatebox{-90}{\epsfbox{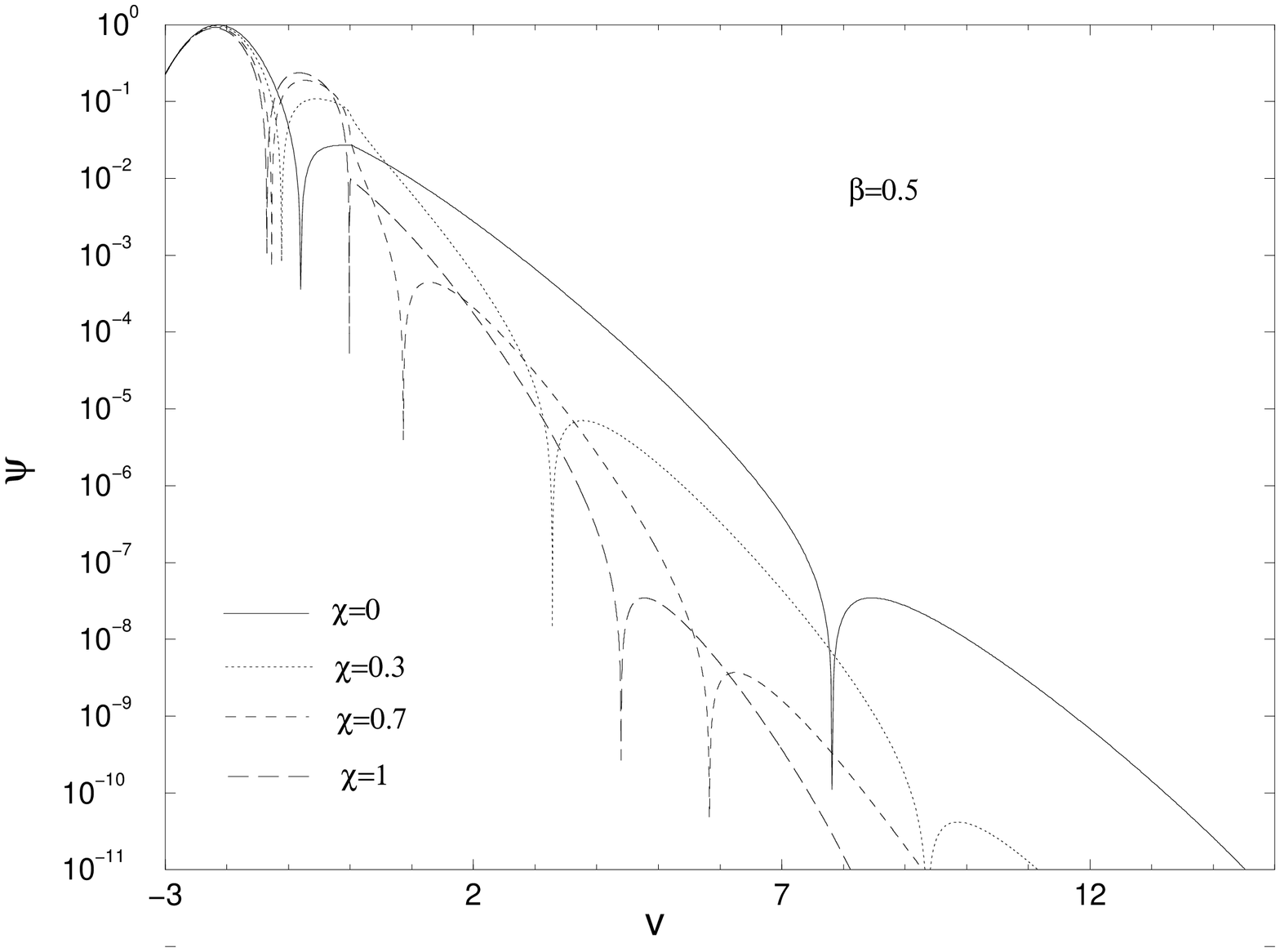}} & & \epsfxsize%
=6truecm\rotatebox{-90}{\epsfbox{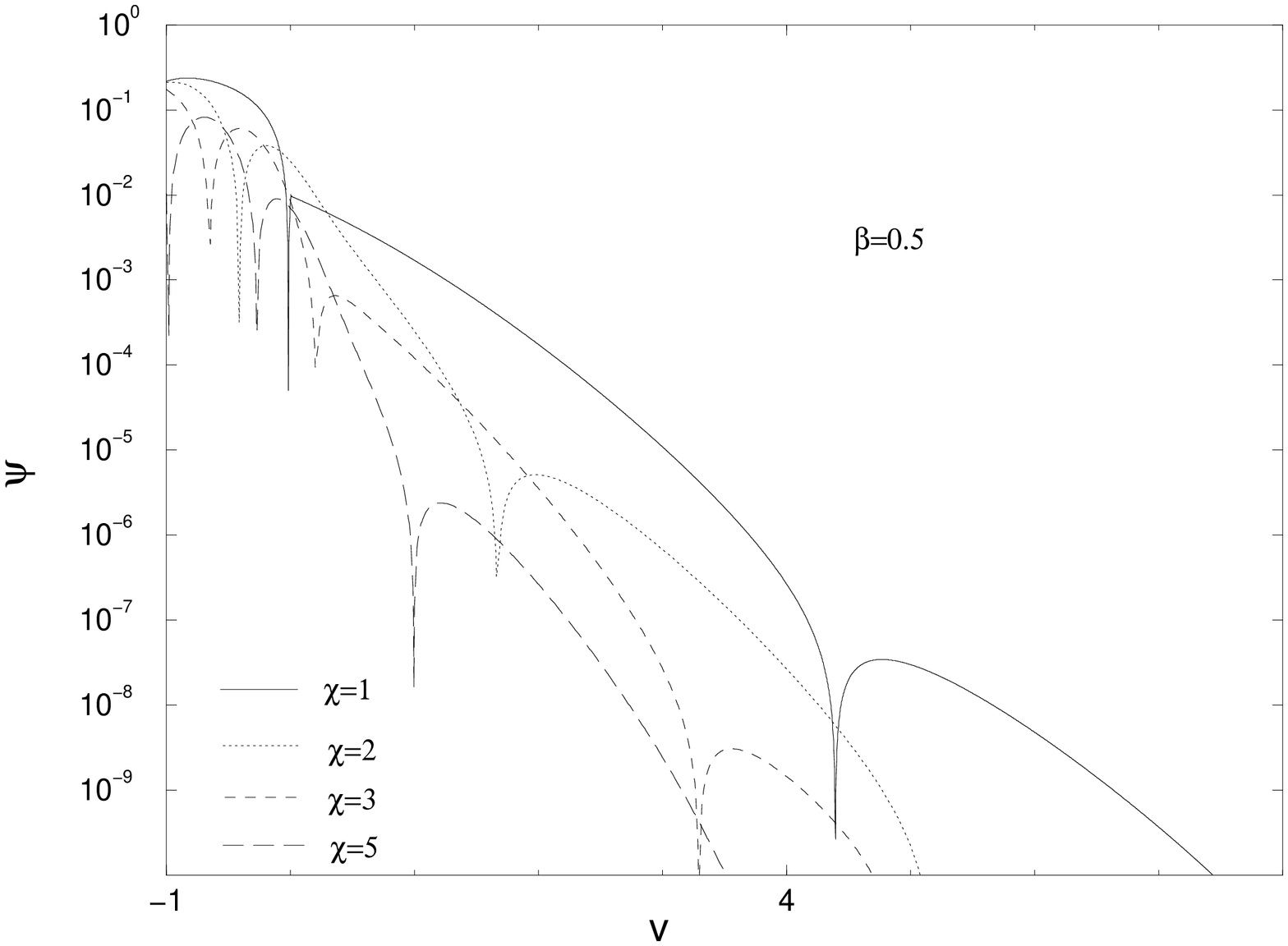}}  \nonumber
\end{eqnarray}
\vskip .5cm
\end{center}
\caption{{Falloff behavior for $\protect\xi \leq 1/6$ in positive mass
topology black holes when $\protect\beta =0.5$}}
\label{fig12}
\end{figure}
\begin{figure}[tbh]
\begin{center}
\leavevmode       
\begin{eqnarray}
\epsfxsize= 8truecm\rotatebox{-90}{\epsfbox{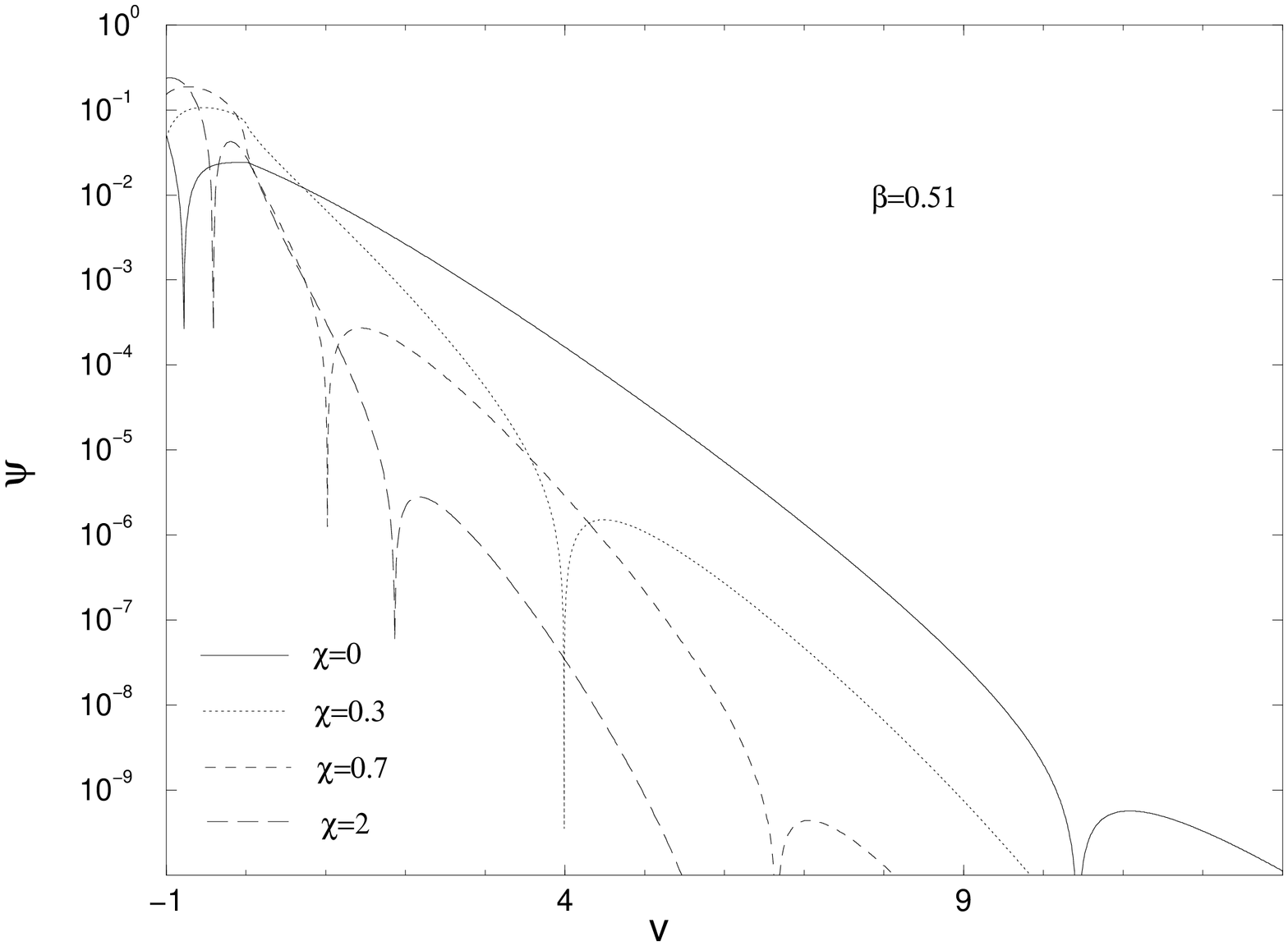}}  \nonumber
\end{eqnarray}
\vskip .5cm
\end{center}
\caption{{Falloff behavior for $\protect\xi \leq 1/6$ in positive mass
topology black holes when $\protect\beta =0.51$ }}
\label{fig13}
\end{figure}
\begin{figure}[tbh]
\begin{center}
\leavevmode       
\begin{eqnarray}
\epsfxsize= 6truecm\rotatebox{-90}{\epsfbox{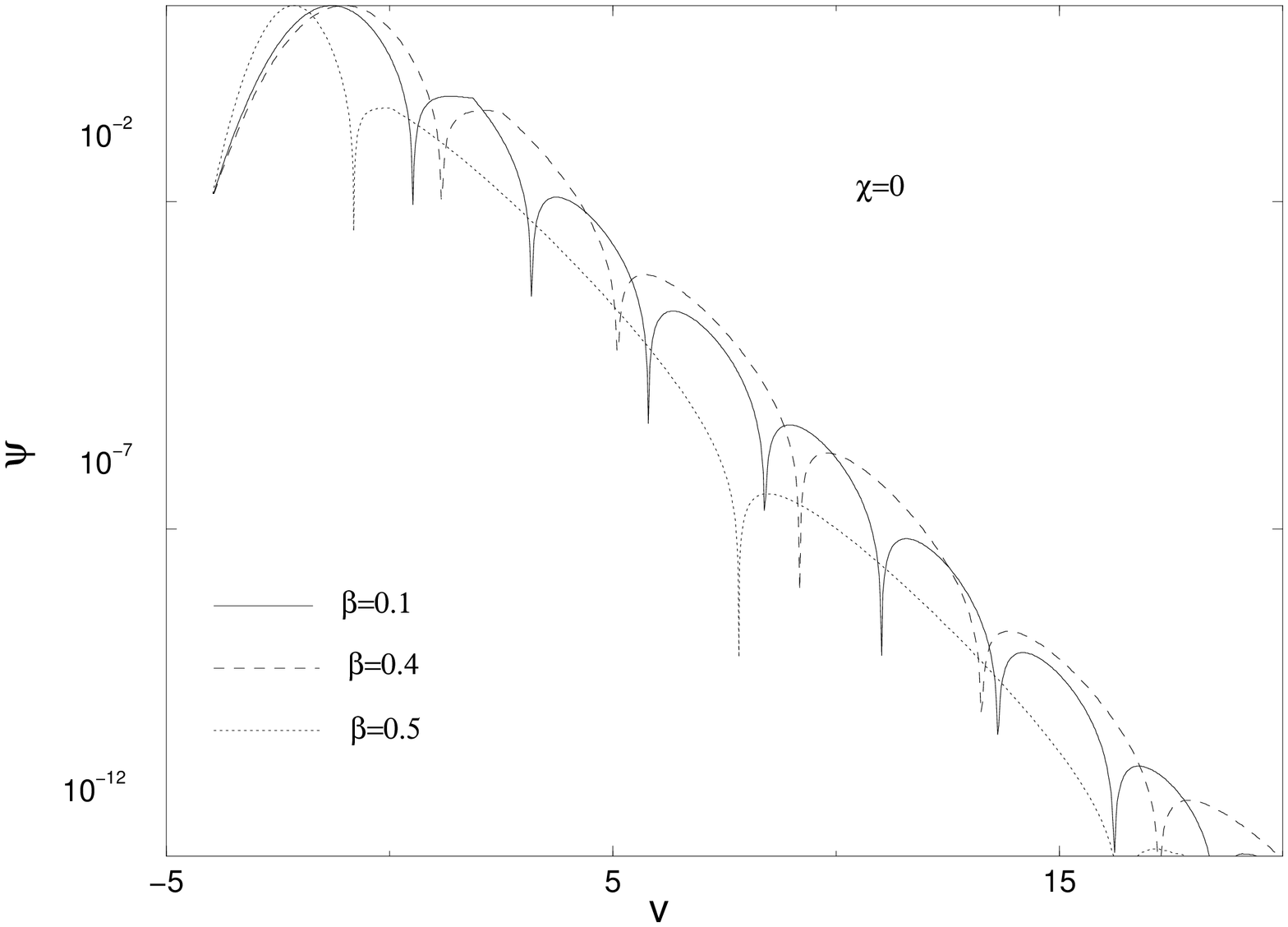}} & & \epsfxsize%
=6truecm\rotatebox{-90}{\epsfbox{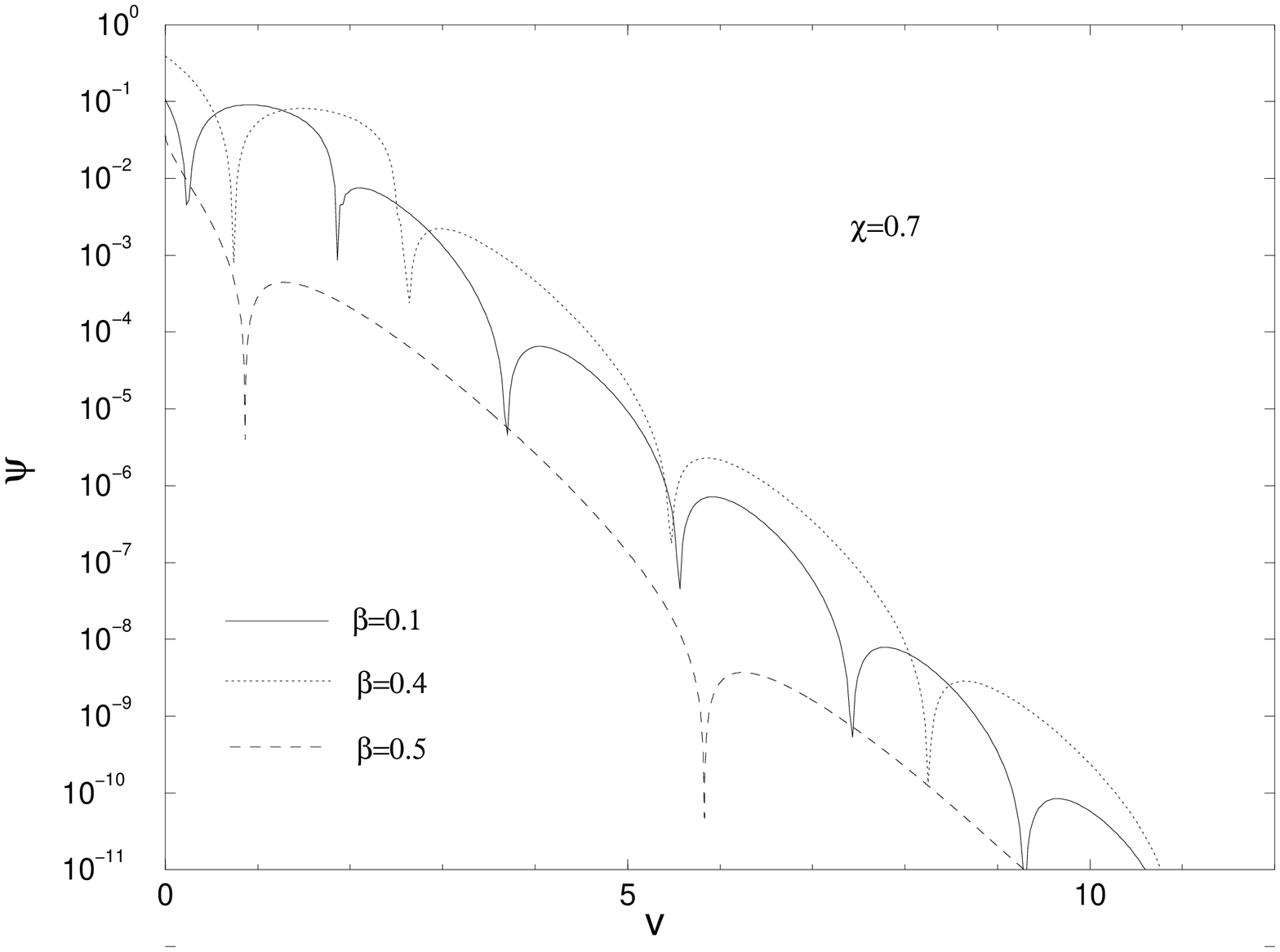}}  \nonumber
\end{eqnarray}
\vskip .5cm
\end{center}
\caption{{wave bahavior with the increase of $\protect\beta $ in positive
mass topological black holes}}
\label{fig14}
\end{figure}
\begin{figure}[tbh]
\begin{center}
\leavevmode       
\begin{eqnarray}
\epsfxsize= 8truecm\rotatebox{-90}{\epsfbox{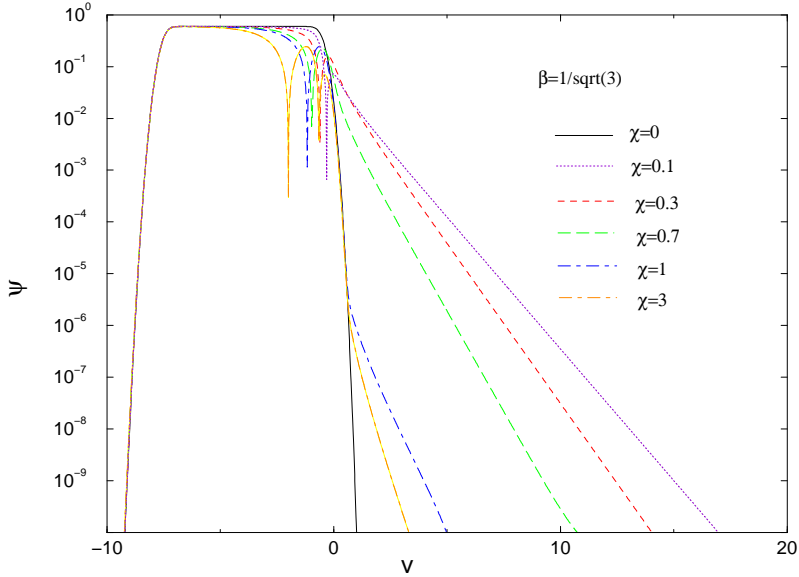}}  \nonumber
\end{eqnarray}
\vskip .5cm
\end{center}
\caption{{Falloff behavior for $\protect\xi \leq 1/6$ in zero mass topology
black holes }}
\label{fig15}
\end{figure}

\begin{figure}[tbh]
\begin{center}
\leavevmode       
\begin{eqnarray}
\epsfxsize= 6truecm\rotatebox{-90}{\epsfbox{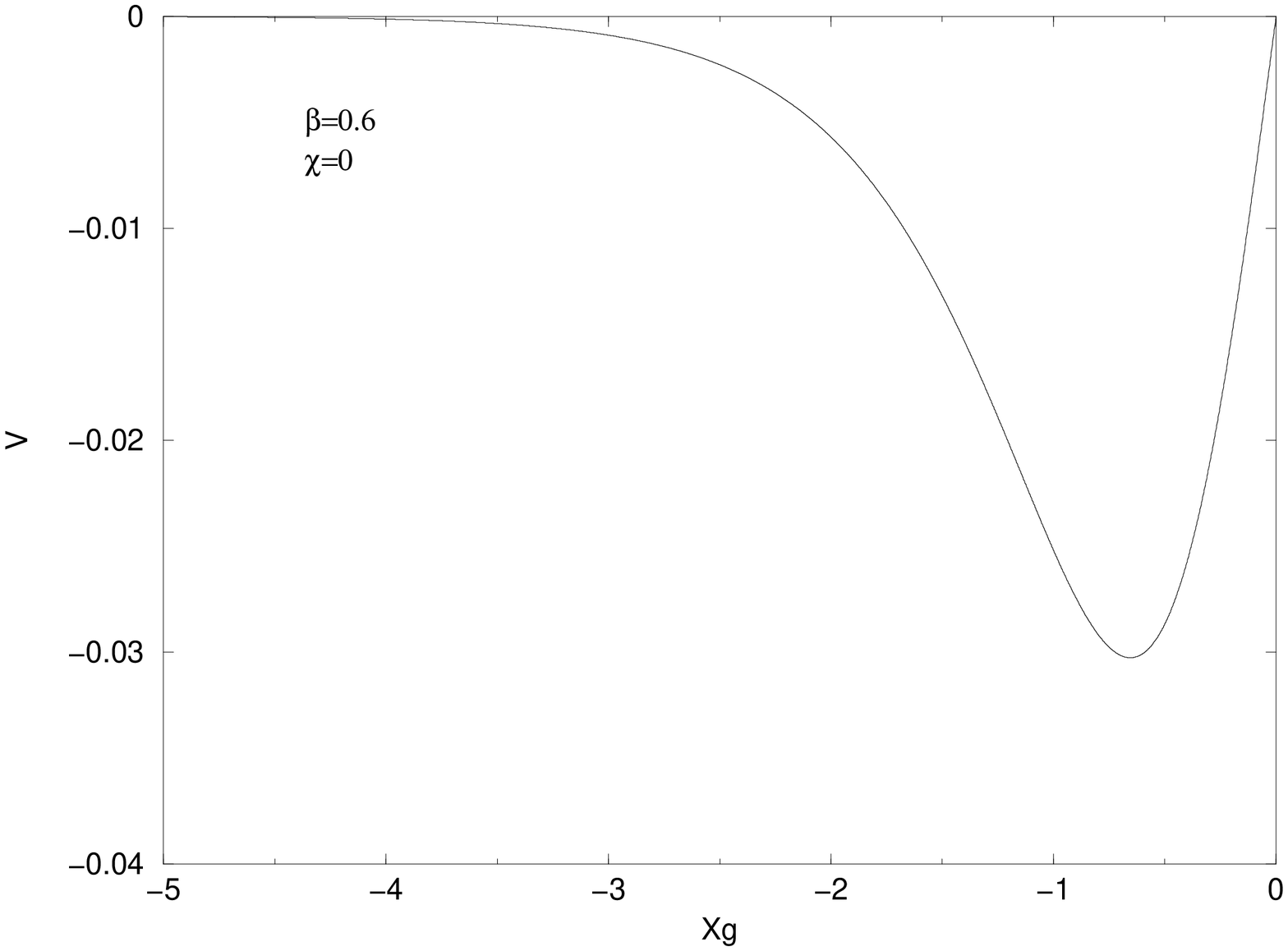}} & & \epsfxsize%
=6truecm\rotatebox{-90}{\epsfbox{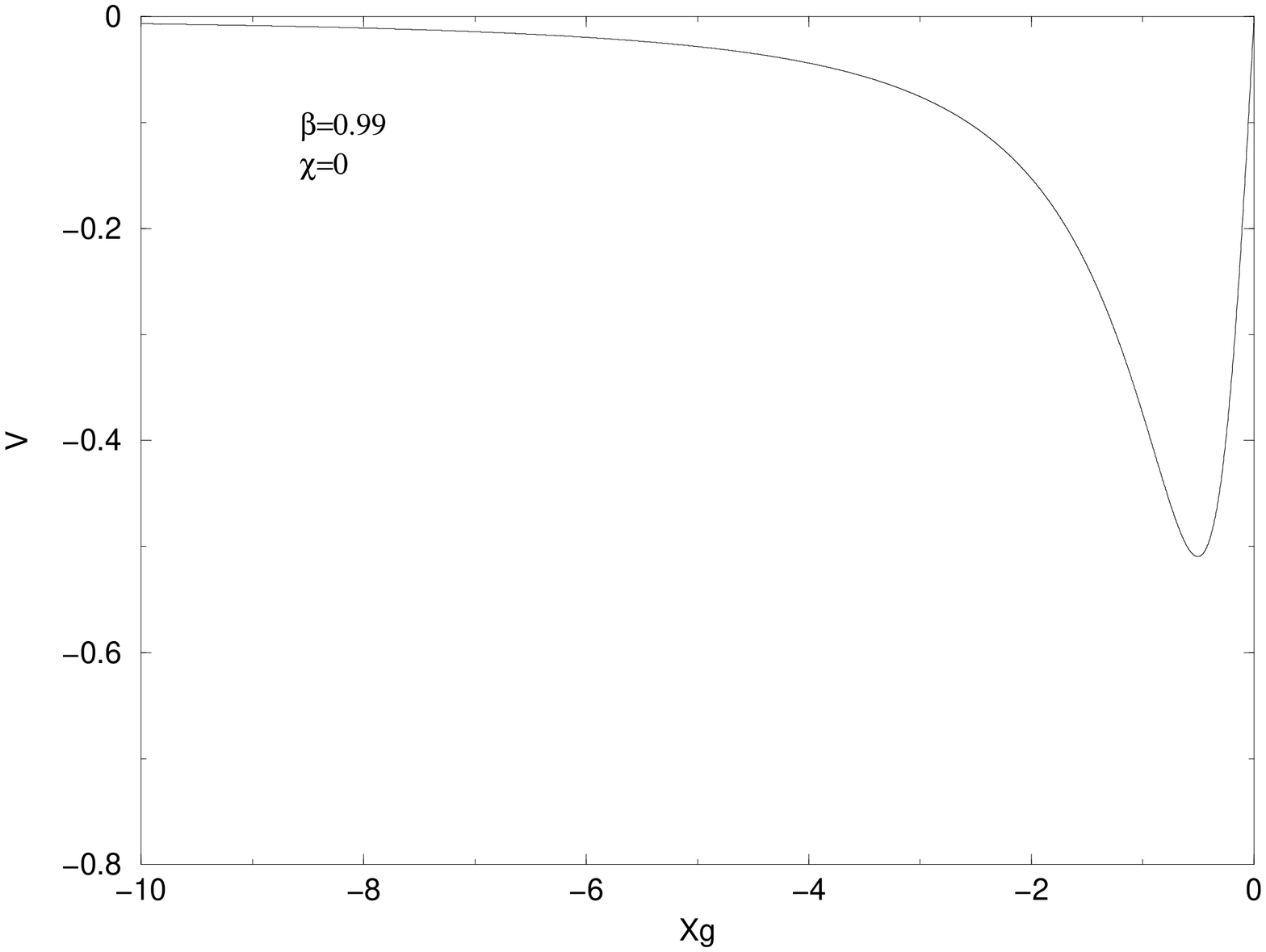}}  \nonumber
\end{eqnarray}
\vskip .5cm
\end{center}
\caption{{Potential behavior for negative mass topology black hole when $%
\protect\xi =1/6$}}
\label{fig16}
\end{figure}
\begin{figure}[tbh]
\begin{center}
\leavevmode       
\begin{eqnarray}
\epsfxsize= 6truecm\rotatebox{-90}{\epsfbox{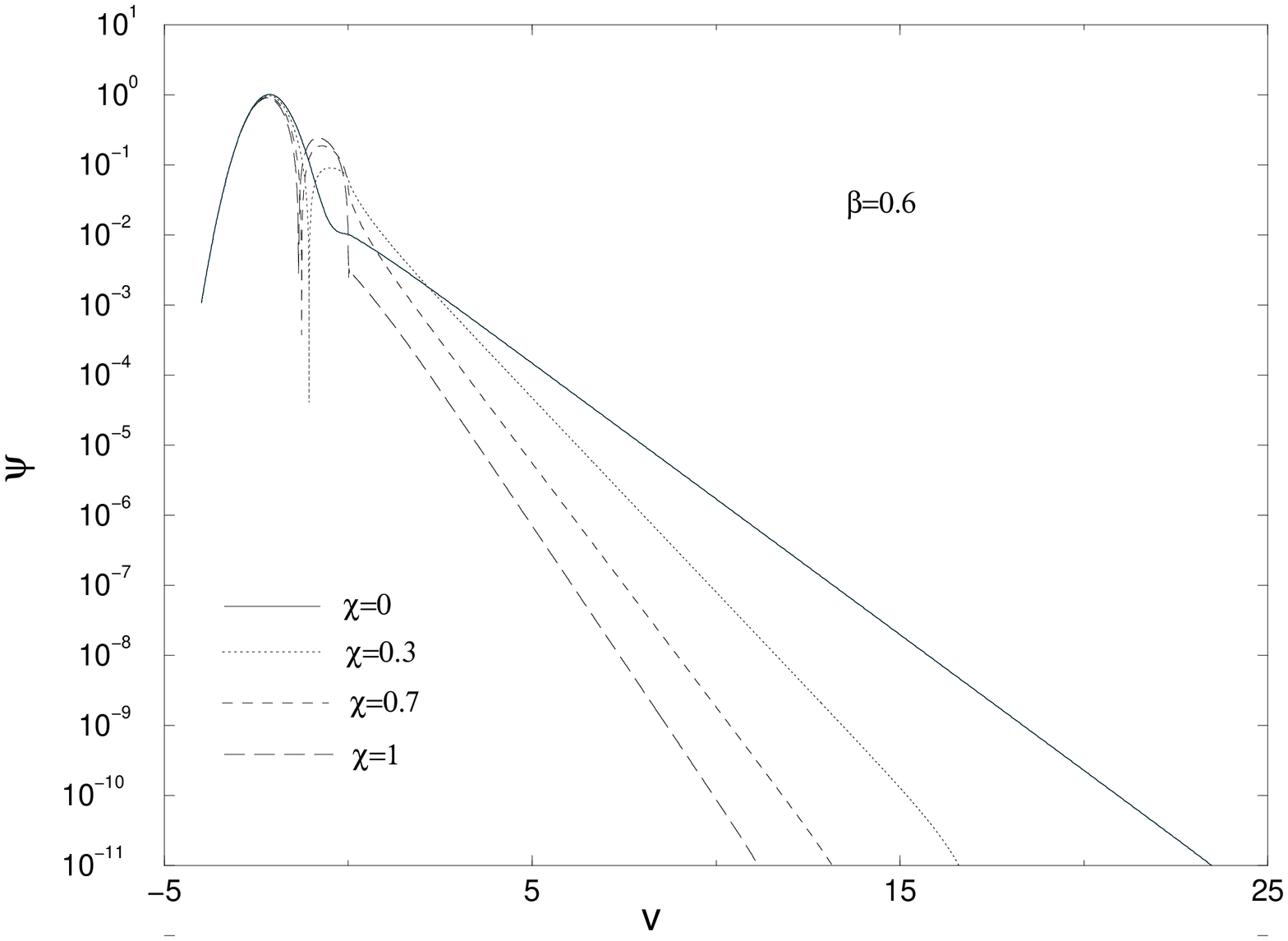}} & & \epsfxsize%
=6truecm\rotatebox{-90}{\epsfbox{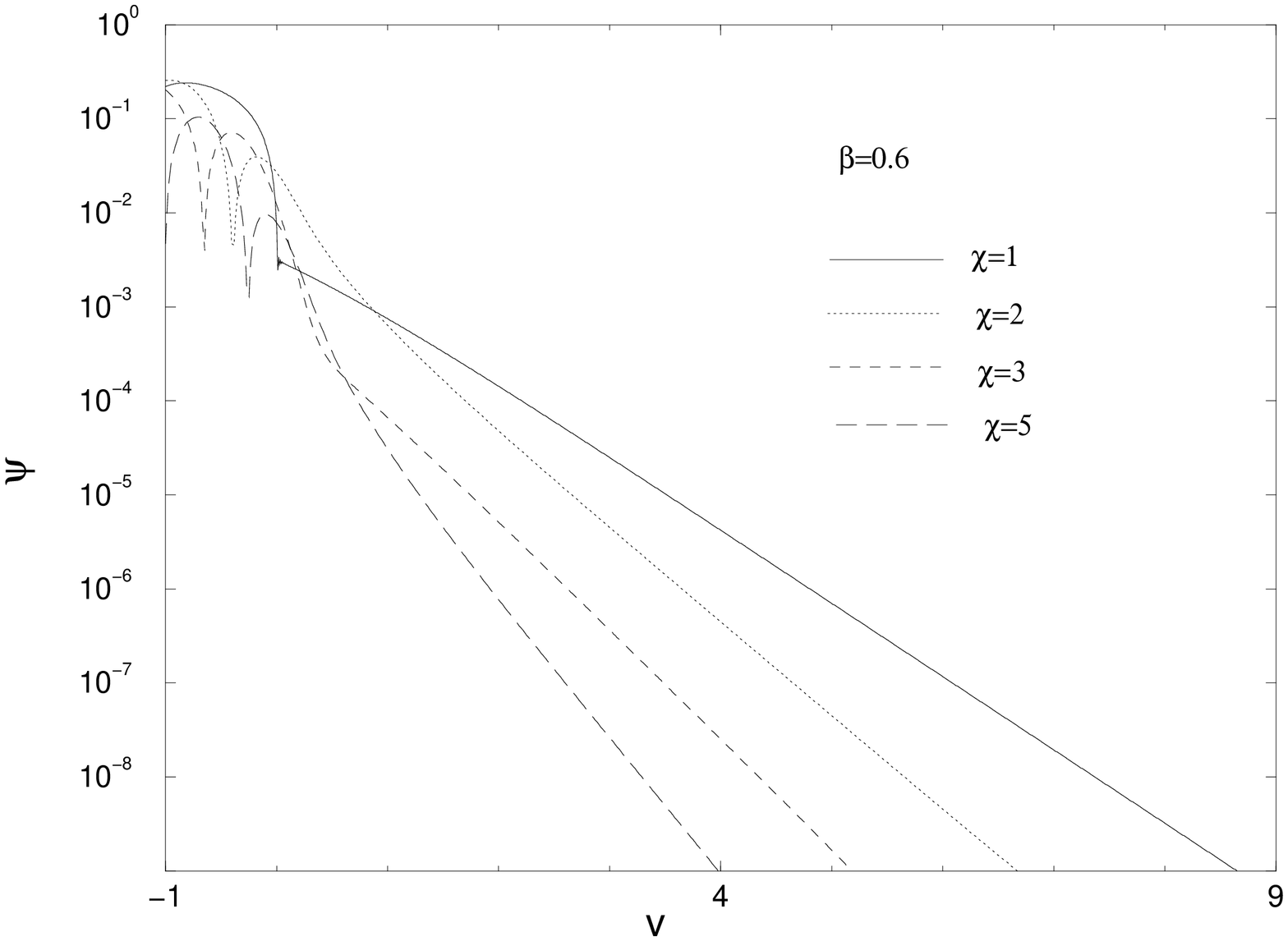}}  \nonumber
\end{eqnarray}
\vskip .5cm
\end{center}
\caption{{\ wave propagation in the negative mass topological black hole
background with different $\protect\xi $ when $\protect\beta =0.6$}}
\label{fig17}
\end{figure}
\begin{figure}[tbh]
\begin{center}
\leavevmode       
\begin{eqnarray}
\epsfxsize= 8truecm\rotatebox{-90}{\epsfbox{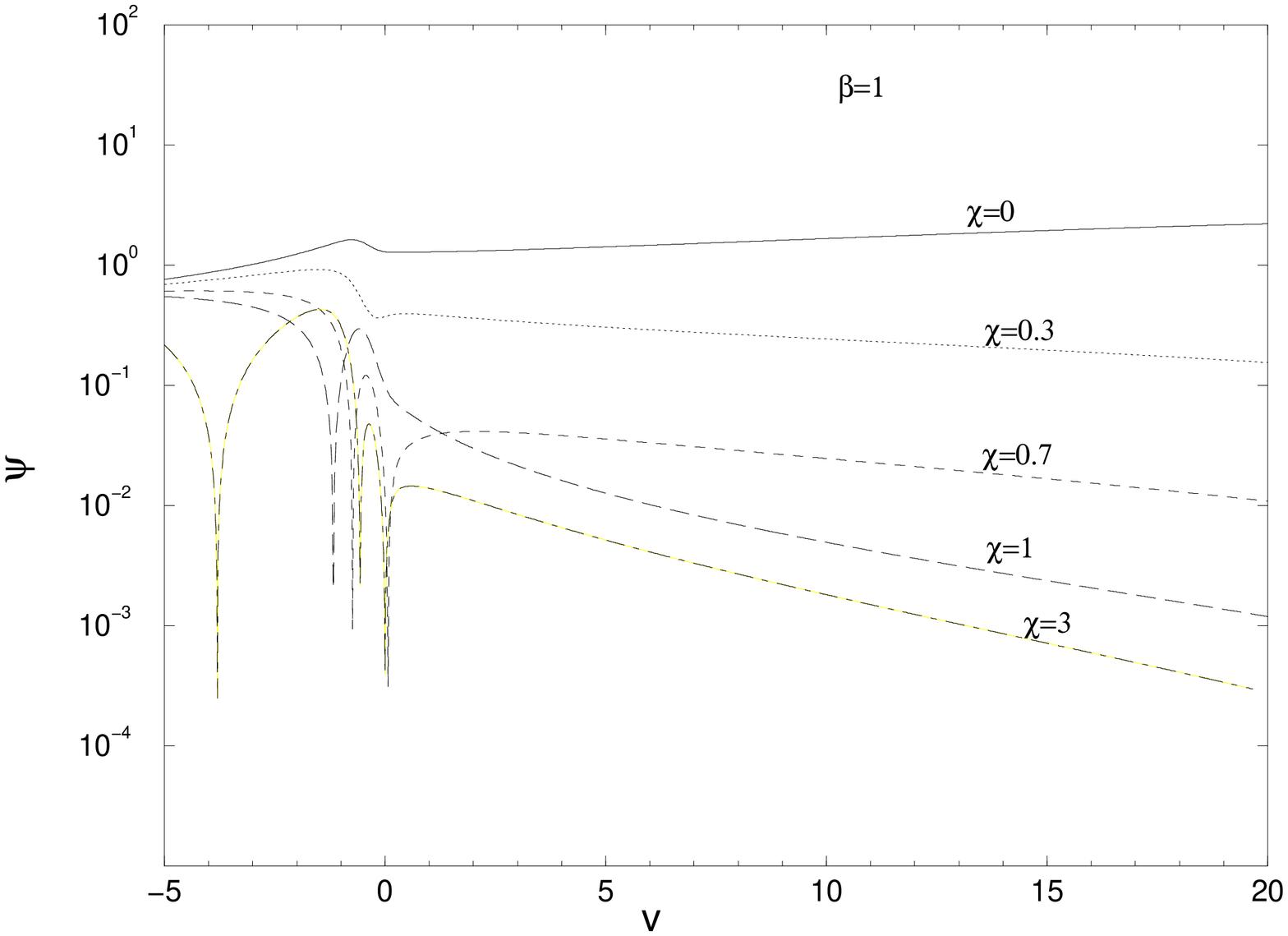}}  \nonumber
\end{eqnarray}
\vskip .5cm
\end{center}
\caption{{wave propagation in the negative mass topological black hole
background with different $\protect\xi $ when $\protect\beta =1.0$}}
\label{fig18}
\end{figure}

\section{Numerical results}

We now report on the results of our numerical simulations of evolving a
massless non-minimally coupled scalar field on different topological black
hole spacetimes. In particulars, we will focus on the dependence of the wave
behaviour on the coupling constant $\xi $. We will show that this dependence
is quite different from that found in de Sitter spacetime \cite{l-p}. We
will also consider the dependence of wave propagation on the parameter $%
\beta $. For the $g=0$ SAdS case, our results are consistent with that of
refs. \cite{horo,wang1} for either large or small holes. For higher genus
holes, we discover some new properties.

\subsection{Schwarzschild AdS black hole $(g=0)$ background}

In the $g=0$ or Schwarzschild AdS case, numerical solutions to wave equation
of scalar field conformally coupled \cite{mann1,mann2} and minimally coupled
to spacetime curvature \cite{horo,wang1} have been obtained. It was found
that the radiative tails associated with a massless scalar wave propagation
have an oscillatory exponential decay. This falloff behaviour depends
neither on the observation point nor on the initial pulse.

In \cite{l-p}, a scalar non-minimally coupled to curvature was discovered to
have a rich spectrum of late-time behaviour in Schwarzschild de Sitter
spacetime. We consider in this section the analogous in SAdS spacetimes. It
is clear from eq. (\ref{Vhat}) that the curvature-coupling constant $\xi =%
\frac{\left( 1-\chi \right) }{6}$ plays an important role in determining the
behaviour of the effective potential ${\tilde{V}}$. For the scalar field
conformally coupled to curvature $\xi =1/6$ $(\chi =0)$, and the graph of
this potential function is given in fig.\ref{fig1}. For $\xi <1/6$ $(\chi
>0) $, the behaviour of the potential is shown in Fig.\ref{fig2}, whereas
fig.\ref{fig3} is the potential for $\xi >1/6(\chi <0)$. The dependence of
scalar wave propagation on the form of the potential has been noted
previously in the asymptotically flat case \cite{ching} .

Figs.\ref{fig4}--\ref{fig7} display the quasinormal ringing of the scalar
field for several values of $\chi (\xi )$ with fixed $\beta $. For
conformally coupled scalar field ($\xi =1/6$), our result coincides with
that obtained in \cite{mann1}. For $\xi \leq 1/6$ $(\chi \geq 0)$, we find
that both the real and imaginary parts of the quasinormal frequencies
increase as $\chi $ increases. This behaviour remains the same as $\beta $
is varied, and differs substantively from that observed in de Sitter
spacetime. Since the quasinormal frequencies of black holes in AdS
spacetimes have a direct interpretation in terms of the dual CFT,
according to the AdS/CFT correspondence the more weakly the scalar field
is non-minimally coupled to the curvature, the faster the thermal state in
the CFT settles down to thermal equilibrium.

%
%

For $\xi >1/6$ , the wave propagation amplifies instead of decays outside
the black hole. This behaviour can be attributed to the negative infinite
potential shown in fig. \ref{fig3}, which implies that the wave outside the
black hole gains energy from the spacetime. The close relationship between
the varying forms of the potential and the rich spectrum of $\xi $%
-dependent\ wave propagation is commensurate with the arguments of Ching et.
al.'s \cite{ching}.

Fig.\ref{fig8} demonstrates the behaviour of the field with increasing $%
\beta $ from $0.4$ to $1$ for $\chi =0$ $(\xi =1/6)$ in the SAdS black hole
background: as $\beta $ increases, the frequency of quasinormal ringing
increases and the rate of decay slows. This qualitative behaviour persists
at other values of $\chi $. From eq. (\ref{eq14}) we see that increasing $%
\beta $ corresponds to a decreasing black hole size according to the
definition given in \cite{horo}. This stands in apparent contrast to that
previously observed for intermediate to small black holes, for which
quasinormal modes decay slower for smaller AdS black holes whilst the
ringing frequency remains nearly the same \cite{horo,wang1}. This apparent
difference is due to the coordinate transformations (\ref{eq11},\ref{eq12})
we employ, which render our time scale different from that used in refs. %
\cite{horo,wang1}. Considering (\ref{eq11}), the relationship between our
frequency ${\tilde{\omega}}$ and the frequency $\omega $ used in refs. \cite%
{horo,wang1} is 
\begin{equation}
{\tilde{\omega}}=\displaystyle\frac{l^{2}}{r_{+}}\omega .  \label{freqrel}
\end{equation}%
From (\ref{freqrel}) we see that an increasing real part of the frequency ${%
\tilde{\omega}}_{R}$ corresponds to a nearly constant value of $\omega _{R}$
as the black hole size decreases, as illustrated in fig.\ref{fig8}.
Likewise, the imaginary part ${\tilde{\omega}}_{I}$ of the frequency will
decrease as $\omega _{I}$ decreases, yielding commensurate decay rates.

%

For big black holes ($r_{+}$ large, $\beta $ very small), we have
investigated $\beta =0.01,0.005,0.001$ etc. using our rescaled coordinates.
We find no difference in wave propagation for these different values of $%
\beta $. From eq.(\ref{freqrel}), this corresponds to an increase of both $%
\omega _{R}$ and $\omega _{I}$ for increasing black hole size, in agreement
with the results of large black holes argued in \cite{horo}. Throughout, we
find that the $\beta $-dependence of the decay rates and frequencies is
qualitatively unchanged for different values of $\xi $.

\subsection{Toroidal black hole $(g=1)$ background}

Now we proceed to discuss the wave propagation in toroidal black hole
background. From (\ref{eq14}) and (\ref{eq16}) we note that the potential
and tortoise coordinate are independent of $\beta $. Consequently the scalar
wave behaves the same outside the toroidal black hole for different values
of $M/l$. The $\xi $-dependence of the potential for toroidal black holes is
qualitatively the same as for the SAdS case, shown in figs.\ref{fig1}--\ref%
{fig3}.

Results displayed in fig.\ref{fig9} illustrate the falloff behaviour for $%
\chi \geq 0$ $(\xi \leq 1/6)$ for toroidal holes. \ It is qualitatively the
same as the SAdS case, although quantitatively the decay rate is larger and
the wave oscillates with higher frequency. In the context of the AdS/CFT
correspondence, any thermal perturbation returns to equilibrium more quickly
outside a toroidal hole as $\xi $ decreases relative to the SAdS case. For $%
\chi <0$ $(\xi >1/6)$, we again find amplification modes.

Similar wave behaviour shown for toroidal holes to Schwarzschild AdS holes
can be attributed to their similar potential properties. When one considers
the argument of Ching et al \cite{ching}, this similarity is not surprising.

\subsection{Higher genus topological black holes ($g\geq 2$)}

When the genus $g\geq 2$, the geometry of the topological black hole depends
on the parameter $\beta $ which runs from $0$ to $1$. There are two distinct
regions of $\beta $ corresponding to qualitatively different black hole
structures. If $0\leq \beta <1/\sqrt{3}$, the black hole has positive mass,
whereas if $1/\sqrt{3}<\beta \leq 1$, the black hole has negative mass. $%
\beta =1/\sqrt{3}$ is associated with a black hole of zero mass.

\bigskip

{\bf A. Positive mass ($0<\beta<1/\sqrt{3}$)}

We solve the scalar wave equation (\ref{eq16}) where the potential $V$ has a
shape similar to the $g=0$ case shown in figs.\ref{fig1}-\ref{fig3}. This
similarity yields wave propagation for different values of $\xi $\ similar
to that of the SAdS case. The results are illustrated in figs.\ref{fig10}-%
\ref{fig13} for fixed $\beta $. We found that for increasing $\chi $
(decreasing $\xi $ in the range $\xi \leq 1/6$), both the real and imaginary
parts of the quasinormal frequency outside the black hole increase. We
thus see that outside the topological black hole with positive mass, if
the field is more weakly non-minimally coupled to spacetime curvature, the
thermal perturbation will settle down faster. Therefore, we qualitatively
observe the same behaviour as that seen in black holes with genus $g=0$
and $g=1$ cases. 


For $\xi >1/6$, the similar divergence of the potential to negative infinity
as shown in fig.\ref{fig3} leads to amplification of the mode again.

Fig.\ref{fig14} illustrates the oscillation behaviour for various values of $%
\beta $. We see that the decay rate (the imaginary part ${\tilde{\omega}}%
_{I} $ of the quasinormal frequency) does not change much as $\beta $
increases. Using the relation (\ref{freqrel}), we see that this property
corresponds to the decrease of both the real $\omega _{R}$ and imaginary $%
\omega _{I}$ parts of the quasinormal frequency with decreasing black hole
size. The range of $\beta $ shown in fig.\ref{fig14} corresponds to
intermediate-size black holes. Compared to intemediate Schwarzschild AdS
holes, we find that falloff behaviour remains similar, while the field
oscillates less rapidly. As $\beta \rightarrow 1/\sqrt{3}$, the ringing
frequency markedly decreases and finally disappears.

For large black holes ($\beta $ very small), no explicit difference has been
observed in wave propagation with different small $\beta $ in our scale. 
This behaviour is similar to $g=0$ case which corresponds to the increase
of both real and imaginary parts of frequencies with the increase of the
size of big black hole in the real time scale.

\bigskip

{\bf B. Zero mass ($\beta=1/\sqrt{3}$)}

According to Eq.(\ref{Vhat}), when $\beta =1/\sqrt{3}$, $V$ vanishes for a
scalar field conformally coupled to curvature for the topological black hole
with genus $g\geq 2$. As a result, the conformally invariant scalar field
propagates freely in topological black hole spacetimes of $\beta =1/\sqrt{3}$%
. The behaviour of the scalar wave conformally coupled to curvature is
illustrated in fig.\ref{fig15} with $\chi =0$, which coincides with that
shown in ref. \cite{mann1}. The ``blip'' in the wave is due to the vanishing
potential and the Dirichlet boundary condition we have taken is as explained
in ref. \cite{mann1}.

It is interesting to study the $\xi $-dependence of wave function. For $\xi
<1/6$, the potential deviates from zero and exhibits behaviour similar to
that shown in fig.\ref{fig2} for the $g=0$ case with increasing of $\chi $
(decreasing $\xi $). The behaviour of the wave for $\xi <1/6$ is shown in
fig.\ref{fig15} for $\chi \neq 0$. We find that the falloff rate increases
with the increasing of $\chi $, in agreement with the aforementioned
behaviour. There is no oscillation of the wave appear here.

For $\xi >1/6$, the potential tends to negative infinity as shown in Fig.\ref%
{fig3}, and amplification appears again.

\bigskip

{\bf C. Negative mass ($1/\sqrt{3}<\beta\leq 1$)}

When $\beta >1/\sqrt{3}$, the black hole mass is negative. This rather
strange situation can be realized by the collapse of matter that violates
the standard energy conditions under certain circumstances \cite{mann4}. For 
$\xi =1/6$ , the potential for $\beta >1/\sqrt{3}$ outside the black hole
becomes everywhere negative. This is quite different from the $g=0,1$ cases
and also different from the positive mass higher genus black hole case. Some
graphs of the potential for $\beta >1/\sqrt{3}$ are shown in Fig.\ref{fig16}%
. We see that larger values of $\beta $ lead to more negative values of the
potential. Combing this property with the conjecture of ref. \cite{horo} in
determining the sign of imaginary part of the frequency (eq.(2.22) in \cite%
{horo}), 
\begin{equation}
\int_{r_{+}}^{\infty }dr[f|\psi ^{\prime }|^{2}+V|\psi |^{2}]=-\displaystyle%
\frac{|\omega |^{2}|\psi (r_{+})|^{2}}{\omega _{I}}  \label{hororel}
\end{equation}%
it might be possible to understand the amplification results shown in \cite%
{mann1} and also in fig.\ref{fig18} below for large $\beta $ when $\xi
=1/6(\chi =0)$. The negative potential for larger $\beta $ changes the sign
of the left-hand-side of the above equation, thereby reversing the sign of $%
\omega _{I}$, yielding amplification.

%

When $\xi <1/6$, the potential becomes positive and tends to infinite at
infinity. The $\xi $-dependence of the potential for $\xi <1/6$ has the same
character as that shown in fig.\ref{fig2}, which contributes to the same
behaviour of the wave for increasing $\chi $ (decreasing $\xi $) in the
range $\xi <1/6$. The scalar fields decays faster the non-minimal coupling
to the spacetime curvature is weaker. This behaviour is shown in figs.\ref%
{fig17},\ref{fig18}.

When $\xi >1/6(\chi <0)$, the potential goes to negative infinity for all $%
\beta $ as shown in Fig.\ref{fig3}. The wave outside the black hole
experiences amplification instead of decay. This result is the same as for
other black holes with $\xi >1/6$.

Comparing figs.\ref{fig17},\ref{fig18}, we find that for fixed $\xi <1/6$,
the wave decays slows with increasing $\beta $. Using eq.(\ref{freqrel}),
this corresponds to the fact that the imaginary part $\omega _{I}$ of
thefrequency decreases with decreasing small black hole size, similar to the
behaviour observed for small SAdS and positive-mass topological black holes.
This agreement is due to the similar behaviour of the potential outside
these various black holes.

\section{Conclusions and discussion}

We have studied the propagation of scalar waves non-minimally coupled to
curvature in black hole backgrounds of different genus $g$. The coupling
constant $\xi $ plays an important role and leads to a rich spectrum of wave
evolution. Physically, non-minimal curvature coupling tends to erode falloff
behaviour: for $\xi <1/6$, the field decays monotonically with a decay
constant that increases with decreasing $\xi $. This behaviour holds for all
topological black holes with differing genus, and is a result of the
qualitatively similar behaviour of their potentials. Our results differ from
those obtained in de Sitter spacetime \cite{l-p} where it was found that the
decay constant increases with increasing $\xi $ for $\xi <1/6$. However in
the de Sitter case the potential falls off exponentially at both the black
hole and cosmological horizons for arbitrary $\xi $. This differs completely
from the situation we are studying here with $\xi <1/6$, and in light of the
observations of Ching et al \cite{ching} not surprising.

The waves outside the black hole experience oscillation for all positive
mass black holes with different topologies. The oscillation frequency
increases as $\xi $ decreases and $\xi <1/6$. For zero mass and negative
mass higher genus black holes, the wave outside the black hole displays only
decay without ringing.

For $\xi =1/6$, which corresponds conformally coupling to the curvature, our
results coincide with those obtained in \cite{mann1}. The wave amplification
for negative mass topological black holes when $\beta $ is close to unity
can be understood by using eq. (\ref{hororel}).

For $\xi >1/6$, potentials tend to negative infinity for large distance
outside topological black holes. Waves amplify instead of decay outside
black holes of any genus. The amplitude of amplification is stronger than
that found for negative mass topological hole with large $\beta $ at $\xi
=1/6$. This is because the potential for $\xi >1/6$ case is more negative
than that negative mass topological with big $\beta $ at $\xi =1/6$.

We have also explored the dependence of quasinormal ringing on $\beta $.
Upon taking account of the coordinate rescaling in (\ref{eq11},\ref{eq12}),
the imaginary part of the quasinormal frequency decreases with $r_{+}$ for
small or intermediate size black holes, regardless of genus. However the
real part $\omega _{R}$ of the quasinormal frequency has some topological
dependence. For intermediate or small SAdS black holes, $\omega _{R}$ does
not change with the black hole size, whereas for positive mass higher genus
small black holes, $\omega _{R}$ decreases with decreasing black hole size.
When the positive mass black hole is small enough ($\beta \rightarrow 1/%
\sqrt{3}$), the wave oscillations disappear. For large black holes ($\beta $
very small), both $\omega _{R}$, and $\omega _{I}$ increase with increasing
black hole size and this property is the same for black holes with $g=0$ and 
$g\geq 2$ cases.

\subsubsection{ACKNOWLEDGMENT:}

This work was partically supported by Fundac\~ao de Amparo \`a
Pesquisa do Estado de S\~ao Paulo (FAPESP) and Conselho Nacional
de Desenvolvimento Cient\'{\i}fico e Tecnol\'{o}gico (CNPQ). B. Wang would
like to acknowledge helpful discussions on numerical problem with T. Osada
and C. Molina and hospitality given by the Dept. of Physics at the
University of Waterloo. The work of B. Wang was also supported by NNSF of
China.


\begin{thebibliography}{99}
\bibitem{kok-sch} K. D. Kokkotas, B. G. Schmidt, {\it Living Rev. Rel.} 
{\bf 2}, 2 (1999), gr-qc/9909058. 

\bibitem{price} R. H. Price, {\it  Phys. Rev.} 
{\bf D 5}, 2419 (1972); {\bf 5}, 2439 (1972). 

\bibitem{g-p-p} C. Gundlach, R. H. Price and J. Pullin, {\it Phys. Rev.} 
{\bf D 49} , 883
(1994). 

\bibitem{bick} J. Bicak, {\it Gen. Relativ. Gravit.} {\bf 3}, 331 (1972).

\bibitem{hod} S. Hod, {\it  Phys. Rev.} {\bf  D 58}, 104022 (1998); {\bf
    61},  024033(2000); {\bf 61},
064018 (2000).

\bibitem{b-o} L. Barack and A. Ori, {\it   Phys. Rev. Lett.} 
{\bf  82}, 4388 (1999); {\it   Phys. Rev.} {\bf   D 60}, 124005 (1999).

\bibitem{a-g} N. Andersson and K. Glampedakis, {\it   Phys. Rev. Lett.} 
{\bf  84}, 4537
(2000).

\bibitem{brady1} P. R. Brady, C. M. Chambers, W. Krivan and P. Laguna,
 {\it    Phys. Rev.} {\bf   D 55}, 7538 (1997).

\bibitem{brady2} P. R. Brady, C. M. Chambers, W. G. Laarakkers and E.
Poisson, {\it   Phys. Rev.} {\bf   D 60}, 064003 (1999).

\bibitem{l-p} W. G. Laarakkers and E. Poisson, gr-qc/0105016.

\bibitem{mann1} J. S. F. Chan and R. B. Mann, {\it   Phys. Rev.} 
{\bf  D 59}, 064025 (1999).

\bibitem{mann2} J. S. F. Chan and R. B. Mann, {\it   Phys. Rev.} 
{\bf  D 55}, 7546 (1997)

\bibitem{horo} G. T. Horowitz and V. E. Hubeny, {\it   Phys. Rev.} 
{\bf  D 62}, 024027 (2000); G. T. Horowitz, {\it    Class. Quant. Grav.} 
{\bf  17}, 1107 (2000). 

\bibitem{wang1} J. M. Zhu, B. Wang and E. Abdalla, {\it   Phys. Rev.} 
{\bf  D 63}, 124004 (2001).

\bibitem{wang2} B. Wang, C. Y. Lin and E. Abdalla, {\it   Phys. Lett.} 
{\bf  B 481}, 79 (2000), hep-th/0003295.

\bibitem{wang3} B. Wang, C. Molina and E. Abdalla, {\it   Phys. Rev.} 
{\bf  D 63}, 084001 (2001).

\bibitem{lemos} V. Cardoso, J. P. S. Lemos, {\it   Phys. Rev.} 
{\bf  D 63}, 124015 (2001);  gr-qc/0105103.

\bibitem{ele} E. Winstanley, gr-qc/0106032.

\bibitem{mann3} R. B. Mann, {\it   Nucl. Phys.} 
{\bf  B 516}, 357 (1998).

\bibitem{mann4} R. B. Mann,  {\it  Class. Quant. Grav.} 
{\bf  14}, 2927 (1997).

\bibitem{ching} E. S. C. Ching, P. T. Leung, W. M. Suen and K. Young, {\it
    Phys. Rev.} {\bf  D 52}, 2118 (1995); {\it   Phys. Rev. Lett.} 
{\bf  74}, 2414 (1995).
\end{thebibliography}
\end{document}